\documentclass[%
 aip,
 amsmath,amssymb,
 reprint,%
]{revtex4-1}

\usepackage{graphicx}
\usepackage{dcolumn}
\usepackage{bm}

\usepackage[utf8]{inputenc}
\usepackage[T1]{fontenc}
\usepackage{mathptmx}
\usepackage{etoolbox}

\usepackage{xcolor}
\usepackage{multirow}
\usepackage{amsmath}
\usepackage{hyperref}
\usepackage{subfigure}
\usepackage{array}
\usepackage{dcolumn}

\makeatletter
\def\@email#1#2{%
 \endgroup
 \patchcmd{\titleblock@produce}
  {\frontmatter@RRAPformat}
  {\frontmatter@RRAPformat{\produce@RRAP{*#1\href{mailto:#2}{#2}}}\frontmatter@RRAPformat}
  {}{}
}%
\makeatother
\begin{document}

\preprint{AIP/123-QED}

\title[Sample title]{A robust single-pixel particle image velocimetry based on fully convolutional networks with cross-correlation embedded}
\author{Qi Gao}
\affiliation{State Key Laboratory of Fluid Power and Mechatronic Systems, School of Aeronautics and Astronautics, Zhejiang University, Hangzhou 310027, China.}

\author{Hongtao Lin}
\affiliation{MicroVec., Inc, Beijing 100083, China.}

\author{Han Tu}
 \homepage{Electronic mail: hantu@zju.edu.cn.}
\affiliation{State Key Laboratory of Fluid Power and Mechatronic Systems, School of Aeronautics and Astronautics, Zhejiang University, Hangzhou 310027, China.}

\author{Haoran Zhu}
\affiliation{State Key Laboratory of Fluid Power and Mechatronic Systems, School of Aeronautics and Astronautics, Zhejiang University, Hangzhou 310027, China.}

\author{Runjie Wei}
\affiliation{MicroVec., Inc, Beijing 100083, China.}

\author{Guoping Zhang}
\affiliation{China Ship Scientific Research Center, Wuxi 214082, China.}

\author{Xueming Shao}
 \homepage{Electronic mail: mecsxm@zju.edu.cn.}
\affiliation{State Key Laboratory of Fluid Power and Mechatronic Systems, School of Aeronautics and Astronautics, Zhejiang University, Hangzhou 310027, China.}

\date{\today}

\begin{abstract}
Particle image velocimetry (PIV) is essential in experimental fluid dynamics. In the current work, we propose a new velocity field estimation paradigm, which achieves a synergetic combination of the deep learning method and the traditional cross-correlation method. 
Specifically, the deep learning method is used to optimize and correct a coarse velocity guess to achieve a super-resolution calculation. 
And the cross-correlation method provides the initial velocity field based on a coarse correlation with a large interrogation window. 
As a reference, the coarse velocity guess helps with improving the robustness of the proposed algorithm.
This fully convolutional network with embedded cross-correlation is named as CC-FCN.
CC-FCN has two types of input layers, one is for the particle images, and the other is for the initial velocity field calculated using cross-correlation with a coarse resolution.
Firstly, two pyramidal modules extract features of particle images and initial velocity field respectively.
Then the fusion module appropriately fuses these features.
Finally, CC-FCN achieves the super-resolution calculation through a series of deconvolution layers to obtain the single-pixel velocity field.
As the supervised learning strategy is considered, synthetic data sets including ground-truth fluid motions are generated to train the network parameters. 
Synthetic and real experimental PIV data sets are used to test the trained neural network in terms of accuracy, precision, spatial resolution and robustness. 
The test results show that these attributes of CC-FCN are further improved compared with those of other tested PIV algorithms. 
The proposed model could therefore provide competitive and robust estimations for PIV experiments.

\end{abstract}

\maketitle
%

\section{Introduction}
\label{sect1}
%
%
%
%

A wide range of problems in engineering applications and scientific research involve extracting physical displacement fields from data \citep{Adrian1991,LiAI2008,Juan2009,Taylor2011},
such as analyzing the internal flow in a cardiac interventional pump \citep{triep2006,triep2008}, studying wake characteristics of fish for the investigation of bio-inspired swimming mechanism \citep{Tytell2008,Ting2009,Flammang2011,Shen2012}, and studying coherent structures in turbulent flows \citep{wang2019experimental,Wang2021}.
Therefore, relevant velocimetry technologies are becoming increasingly important in the field of experimental fluid mechanics as they help researchers to get further understanding and deeper insight into the complex flow phenomena \citep{cai2019}.
As a non-invasive measurement technique, particle image velocimetry (PIV)\citep{Raffel2007,Adrian2011} has been widely applied to optical, quantitative, and non-contact surface or spatial fluid analysis.
The use of PIV allows the spatial development of the flow field to be investigated in detail \citep{Pratt2013}.
Velocity fields can be obtained by this method. Then related properties of the flow, such as vorticity, acceleration \citep{IPTA} and pressure \citep{irro-pre,PCS}, can be derived from the velocity measurement result.
The basic principles of PIV can be summarized as follows. 
Tracer particles, which are sufficiently small to faithfully follow the flow motions, are uniformly seeded into the flow medium. 
The fluid with entrained particles is illuminated by a laser sheet so that the particles become visible in the measurement region.
Successive images with illuminated particles are recorded by one or multiple cameras, depending on the type of the PIV technique.
Finally, the speed and direction (i.e., the velocity vector field) of the flow in the measurement region are obtained by calculating the motion of the tracer particles recorded in particle images.

In practical applications, what researchers ultimately need is the high-precision and high-resolution velocity field. 
Therefore, how to properly estimate the velocity field of the flow from particle images becomes very important. 
The image analysis technique for calculating the velocity field from particle images has been developed for decades and several different calculation methods have gradually formed.
One of them is the cross-correlation approach, which could provide sparse motion field by searching the maximum of the cross-correlation between two interrogation windows of an image pair \citep{Raffel2007,Adrian2011,Westerweel1999}.
Regarding the displacement vector as the average velocity in the window, the cross-correlation method is widely used in commercial software due to its simplicity and efficiency.
\citet{Scarano2001} proposed a window deformation iterative multi-grid (WIDIM) method, which could optimize the numerical calculation approach and improve the accuracy and resolution of the calculation results. 
The WIDIM method achieves a decoupling between the spatial resolution and the dynamic range by using an iterative evaluation program with integrated window refinement. It achieves good performance evaluation in the International PIV Challenges \citep{Stanislas}.
Another approach to obtaining the velocity vector field from particle images is the optical flow computation, which was first proposed by \citet{Horn1981}.
This method has been very popular in the computer vision community and attracts researchers in experimental fluid mechanics since it gives dense motion field for the whole image. 
At present, a series of different types of optical flow methods have been developed in order to obtain better results in fluid motion analysis, such as \citet{Corpetti2002,Ruhnau2007,Zhong2017,lu2021accurate}.
However, optical flow methods are generally time-consuming as they contain optimization processes to minimize the objective function \citep{cai2019,cai2019particle}. In addition, the brightness constancy constraint in optical flow methods is sensitive to the image noise \citep{cai2019particle}.
Different from the above-mentioned reconstruction methods in the Eulerian description, particle tracking velocimetry (PTV) is a Lagrangian method which tracks the trajectory of particles. 
This approach could potentially offer a higher spatial resolution than PIV provides, since it avoids the spatial filtering effect inherent to the cross-correlation algorithm \citep{Schneiders2016}. 
The downside of PTV is the limitation in particle image density \citep{Schneiders2016,Schanz2016}.
With the development of Shake-The-Box (STB)\citep{Schanz2016} and advanced iterative particle reconstruction (IPR)\citep{Wieneke}, the Lagrangian particle tracking method can have seeding concentrations on a higher order. 

Compared with the manually designed operations, deep learning is a completely different approach to solving problems. 
This method can learn an estimating model which can approximate any desired complex mapping function from data. 
Deep learning has achieved great success in the computer image processing, and it presents a great potential in the field of PIV.
Although PIV deals with the physics-related problem, estimating the motions of particles from images is considered as an issue of image processing or computer vision to some extent.
Hence, designing a deep neural network for the flow motion estimation becomes a promising direction.
\citet{rabault2017} reported the first application of Convolutional Neural Networks (CNNs) and Fully Connected Neural Networks (FCNNs) in performing end-to-end PIV. 
Although the proposed models in \citet{rabault2017} did not outperform the traditional PIV methods with respect to root mean square error, the study aroused attention and inspired other researchers to involve in this topic. 
\citet{lee2017} designed a four-layer cascaded network architecture (PIV-DCNN) to improve the performance and spatial resolution of the large displacement estimation. 
\citet{wang2020} developed a convolutional neural network to measure the velocity field of near-wall turbulence. 
The proposed machine learning based model can improve the near-wall velocity prediction and spatial resolution of the wall turbulence velocity field obtained by PIV.
\citet{dosovitskiy2015} proposed the first networks (FlowNetS and FlowNetC) for optical flow estimation. As the successor, FlowNet2 \citep{Ilg2017} was designed as a cascade of variants of the FlowNet.
\citet{Sun2017} proposed the PWC-Net model for optical flow estimation.
\citet{cai2019} modified the FlowNetS and used it to estimate the dense velocity field of PIV.
Additionally, they also modified the LiteFlowNet \citep{hui2018} convolutional neural network to further improve the estimation performance of the machine learning method \citep{cai2019particle}. 
The original network is improved mainly by adding a deconvolution layer and adjusting the weight of the loss function, so that it can extract a more accurate and precise velocity field from the particle image pair.


Although the PIV technique based on machine learning has shown advantages such as higher accuracy and higher spatial resolution, the stability and robustness of relevant models still need be further improved for practical applications.
Severe noise in particle images brings challenge to the estimation.
Moreover, system complexity and training determine the generalization capability of the network. 
If the network is over-trained or system complexity is more than the training dataset, poor generalization could be observed \citep{Urolagin}. In other words, traditional CNN-based flow motion estimators may give unsatisfactory results when predicting unseen or complicated flow fields in real applications.
In the current work, we propose an innovative deep learning model embedded with cross-correlation, which can suppress noise and obtain satisfying results for practical applications.

The article is organized as follows. In Section~\ref{sect2}, the basic concept and principles of the proposed model are provided.
In Section~\ref{sect3}, we describe the structure of the proposed neural network and the generation of synthetic training data. 
The test results on real and synthetic PIV experimental data are discussed in Section~\ref{sect4}, followed by the conclusions in Section~\ref{sect5}.


\section{Proposed method}
\label{sect2}
In the PIV experiment, in order not to affect the fluid dynamics of the flow being studied, the density of the tracer particles seeded into the flow medium cannot be infinitely high.
Therefore, it is impossible to have tracer particles at every pixel.
Even if there is an ideal method which can obtain the displacement of each tracer particle, it still cannot get the accurate displacement information at the region where there is no particle. 
As a result, PIV cannot obtain the flow information of arbitrary places in the measurement region, but can only obtain the average velocity within the interrogation window.
To sum up, the information about fluid motion recorded by the PIV experiment is filtered. 
This characteristic of PIV fundamentally limits the extraction of single-pixel-level high-resolution flow fields through direct calculation methods, such as matching tracer particles of two adjacent particle images.

Through the above analysis, the PIV experiment can be considered as a down-sampling operation of the flow field, which records the down-sampled flow motion information in the form of particle images. 
Given a real flow field, denoted by $I_{x}$, the output of the PIV experiment carried out on it can be modeled as
\begin{equation}
I_{y}=D(I_{x},\delta)+\eta,
\label{eqn1}
\end{equation}
where $D$ denotes a down-sampling function, $I_{y}$ is the particle image, $I_{x}$ is the ground truth flow field, $\delta$ represents the parameters of the down-sampling process, and $\eta$ is the noise introduced during the PIV experiment. 
Generally, details of the down-sampling process (i.e., $D$ and $\delta$) and the noise $\eta$ are unknown. 
In this case, our goal is to recover an approximation $\hat{I}_{x}$ of the ground truth $I_{x}$ from the particle image $I_{y}$, following:
\begin{equation}
\hat{I}_{x}=F(I_{y},\theta).
\label{eqn2}
\end{equation}
Here $F$ is the reconstruction model and $\theta$ denotes the parameters of $F$. 
The function of $F$ is to recover the entire flow field from $I_{y}$, which records the filtered information of the ground truth flow field.
This is essentially a super-resolution problem, and the noise $\eta$ is an important factor for a successful execution of the calculation.
Improper handling of the noise $\eta$ could lead to calculation failure during the reconstruction process of the deep learning method.
In the process of neural network model training, the actual noise $\eta$ is usually reduced as the Gaussian noise.  
However, in addition to the Gaussian noise, there are many kinds of noise in actual PIV experiments, which vary from experiment to experiment.
Therefore, it is impractical to summarize these different kinds of noise and put them into the training set to search for a well-trained neural network model.
When a new form of noise appears in the particle image, it may cause the calculation to fail.

The key to meeting the above-mentioned challenge is to find a way to effectively suppress the noise $\eta$ in PIV particle images, and let the neural network perform reconstruction calculations based on as much accurate information as possible.
It is well known that the cross-correlation algorithm has strong robustness and can resist the influence of noise in particle images.
The output of the cross-correlation algorithm can be modeled as follows:
\begin{equation}
I_{z}=C(I_{y},\delta_{c})+\eta_{c},
\label{eqn3}
\end{equation}
where $C$ denotes the cross-correlation function, $I_{y}$ is the particle image, $\delta_{c}$ is the parameters of the cross-correlation function and $\eta_{c}$ is the noise error.
$I_{z}$ is the calculated velocity field obtained by the cross-correlation algorithm.
Although its resolution is severely reduced, $I_{z}$ still gives most of the information about the flow field. 
More importantly, $\eta_{c}$ is small enough so that the result of the cross-correlation calculation can be used as the standard reference. 
The role of the cross-correlation algorithm is equivalent to an information converter which filters noise. Specifically, the cross-correlation algorithm converts and down-samples the ground truth information to the form of the sparse velocity field. This procedure filters most of the noise.
Based on the above analysis, a new reconstruction model based on deep learning is proposed as follows:
\begin{equation}
\hat{I}_{x}=F_{c}(\lambda_{1}I_{y},\lambda_{2}I_{z},\theta_{c}),
\label{eqn4}
\end{equation}
which can be written in a more specific form:
\begin{equation}
\hat{I}_{x}=F_{c}(\lambda_{1}D(I_{x},\delta)+\lambda_{1}\eta,\lambda_{2}C(I_{y},\delta_{c})+\lambda_{2}\eta_{c},\theta_{c}).
\label{eqn5}
\end{equation}
Here $F_{c}$ is the reconstruction model based on deep learning proposed in the current work. 
$\lambda_{1}$, $\lambda_{2}$ and $\theta_{c}$ denote the parameters of $F_{c}$, which can be learned automatically in the process of neural network model training.
The new reconstruction model has two different inputs, one is the particle images $I_{y}$, and the other is the velocity field $I_{z}$ calculated by the cross-correlation method. 
In the new reconstruction model, the noise introduced in the PIV experiment becomes $\lambda_{1}\eta$.
Specifically, the neural network model can get a proper parameter $\lambda_{1}$ by learning to suppress the noise.
Then, the noise can be reduced to an appropriate level to avoid calculation failure. 
Although the parameter $\lambda_{1}$ also affects the neural network model to obtain information from the particle images, calculations on actual cases show that the neural network model can extract enough information from the flow field $I_{z}$ with small noise error $\eta_{c}$ to complete the velocity field extraction.

To this end, the objective of the reconstruction model proposed in the current work is to minimize the mean square error (MSE) between the generated flow field $\hat{I}_{x}$ and the ground truth flow field $I_{x}$ as follows:
\begin{equation}
\hat{\theta_{c}}=argmin_{\theta_{c}}L(\hat{I}_{x},I_{x}),
\label{eqn6}
\end{equation}
where $L(\hat{I}_{x},I_{x})$ represents the loss function between $\hat{I}_{x}$ and $I_{x}$. The loss function used in this work is the most popular pixel-wise mean square error.
The $argmin_{\theta_{c}}$ finds the smallest possible value of the loss function $L$ by adjusting $\theta_{c}$. 
And $\hat{\theta_{c}}$ represents the optimal parameters for the reconstruction model $F_{c}$.

\begin{figure*}
\centering
  \includegraphics[width=1\textwidth]{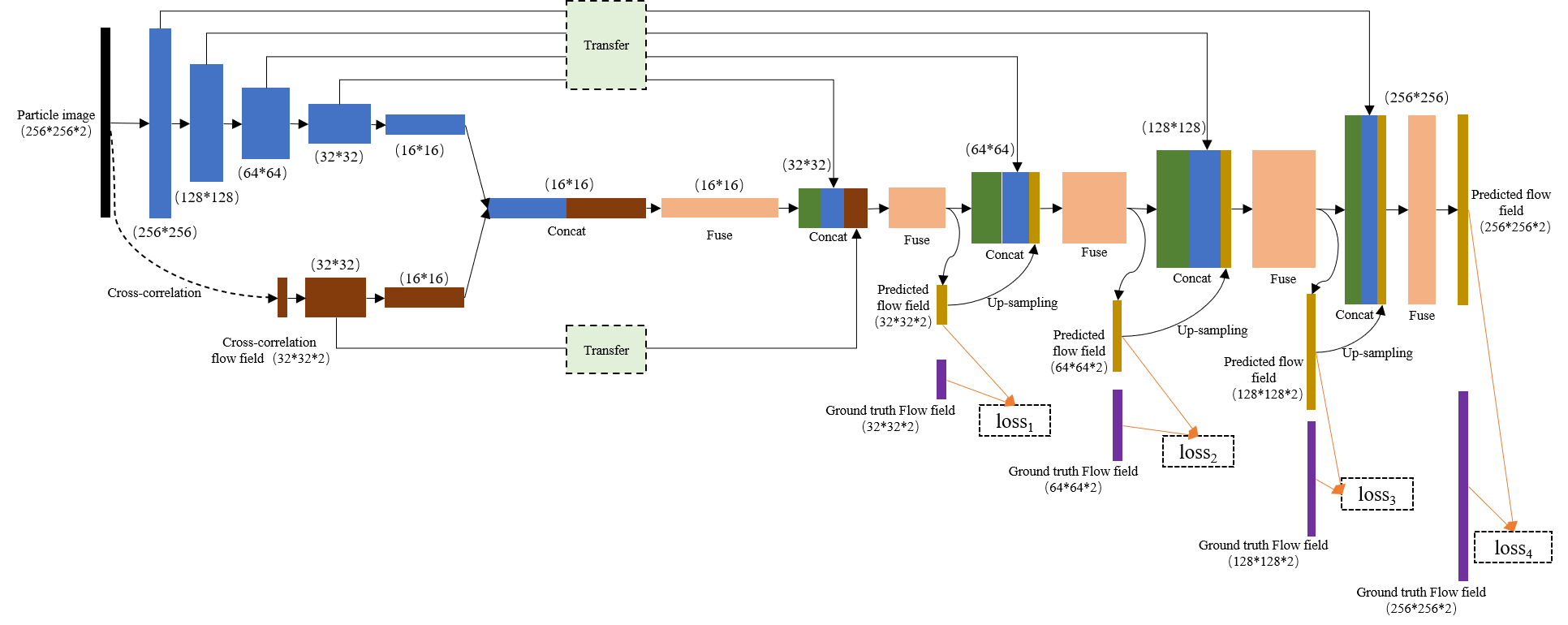}
\caption{Architecture of the CC-FCN model.}
\label{fig1}  
\end{figure*}

Tests on real PIV images show that the reconstruction model proposed in the current work combines the advantages of traditional cross-correlation-based algorithms and purely artificial intelligence (AI) based algorithms. 
It is not only robust to significant noise, but also capable of obtaining high-resolution flow fields with the single-pixel level.

\section{Neural Network design and dataset training}
\label{sect3}
\subsection{Neural Network for PIV}
\label{sect3.1}

The neural network architecture proposed in this paper is based on the Fully Convolutional Networks (FCN)\citep{long2015}.
FCN is a special case of convolutional networks (ConvNets). The training process of FCN is end-to-end (pixel-to-pixel) and embodies less network parameters. 
Thus, it can simplify and speed up the processes of learning and inference in the networks. As a result, the time cost of the network training is reduced while the computational efficiency of the network is improved.  
More importantly, FCN is able to take input of arbitrary size and generate output of the same size.
It can generate dense pixel output via deconvolution layers \citep{nie20183}. 
FCN has achieved state-of-the-art performance in many image processing applications. 
Because of these outstanding features, we employ and further extend FCN, and then propose a new network structure based on fully convolutional networks with cross-correlation embedded, named as CC-FCN.
The structure of CC-FCN is demonstrated in Fig.~\ref{fig1} and a brief introduction is given as below.

As illustrated in Fig.~\ref{fig1}, CC-FCN has two separate input layers, including the image input layer and the velocity input layer. 
Particle images are used as input in the image input layer, and the velocity field calculated by the cross-correlation method with very coarse resolution is used as a reference velocity in the velocity input layer. 
After the two input layers, there are two pyramid-shaped contraction network architectures composed of multiple convolution layers and pooling layers for gradually extracting features from the particle images and the coarse velocity field respectively. 
The kernel of the convolution layers is set to (3,3) and the Leaky linear correction unit (LeakyReLU)\citep{xu2015empirical} is used as the activation function. 
Maximum pooling layers with a step size of 2 are used in the pyramid-shaped contraction network structures. 
Then a concatenate layer is used to merge features extracted from the particle images and the coarse velocity field. 
After the concatenate layer, a convolution layer is used to extract features from the concatenate layer, thereby further enhancing the fusion of the two types of input information.
Finally, an expanding part composed of a number of deconvolution layers is used to gradually up-sample the fused features, and ultimately the velocity field with the same size as the input particle images is obtained.

According to the above introduction, the CC-FCN model consists of three major operations: two down-sampling and one up-sampling. 
The two down-sampling operation streams (i.e., convolution and pooling), extracting information from the particle images and the velocity field calculated by the cross-correlation respectively, usually result in a coarse and global prediction based on the entire input of the network.
And the up-sampling stream (i.e., deconvolution) can generate the dense prediction through finer inference. 
The information reaching up-sampling phases is highly abstracted after a series of pooling operations. Thus, it is easy to lose small-scale structures.
To solve this issue, an information transfer architecture for CC-FCN is designed.
Specifically, we transfer the context information of the coarse feature maps obtained from the set of down-sampling operations to the up-sampling phases, and concatenate these transferred features with features obtained from deconvolution.
The transfer operation not only copies the feature maps, but also uses the convolutional layer in the transfer structure to adjust the number of feature maps from the lower layer to be comparable to the number of those from the corresponding higher layer.
More importantly, we employ fusion modules (i.e., extra convolutional layers) after the concatenation layers to enhance the fusion of the low-level and high-level features. 
As a consequence, the deconvolution layers during the up-sampling phases can generate more precise outputs based on the assembled feature maps.

In order to train the neural network more effectively, we design four output layers to output the predicted velocity fields at different levels. 
On the one hand, the predicted velocity fields are used to construct the loss function. On the other hand, they are up-sampled and concatenated to the next level.
For CC-FCN, the loss function is composed of the prediction errors at different levels of the expanded network architecture:
\begin{equation}
Loss=\Sigma_{i}\lambda_{i}e_{i},
\label{eqn7}
\end{equation}
where $i$ is the level index, $e_{i}$ denotes the error metric between the ground truth and the prediction, and $\lambda_{i}$ represents the preset weight of different levels.

\subsection{Training dataset}
\label{sect3.2}

\begin{figure*}
\begin{center}
  \includegraphics[width=1\textwidth]{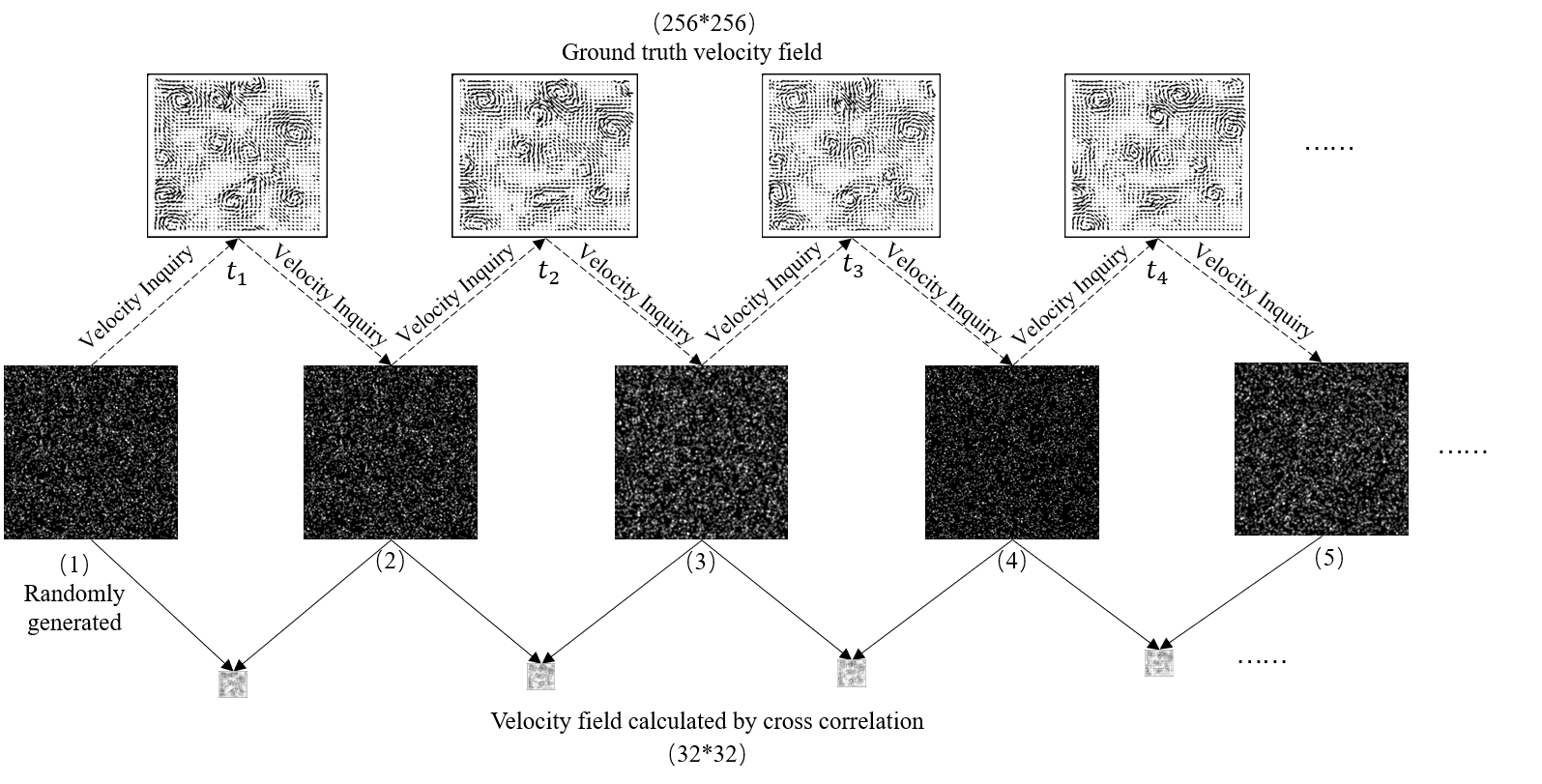}
\caption{Generation of training data sets.}
\label{fig2}   
\end{center}
\end{figure*}

\begin{table*}
\begin{center}
\caption{Descriptions of the motion fields}\label{tab1}%
\begin{tabular}{ p{3cm} p{6cm} p{3cm}  p{2cm} }  
\hline
Case name & Description  & Condition & Quantity\\
\hline
Uniform    & Uniform flow   & $|dx|\in[0;5]$  & 1000  \\
\hline
\multirow{4}*{Back-step} & \multirow{4}*{Backward stepping flow} & $Re=800$ & 600 \\
~ & ~ & $Re=1000$ & 600 \\
~ & ~ & $Re=1200$ & 1000 \\
~ & ~ & $Re=1500$ & 1000 \\
\hline
\multirow{5}*{Cylinder} & \multirow{5}*{Flow over a circular cylinder} & $Re=40$ & 50 \\
~ & ~ & $Re=150$ & 500 \\
~ & ~ & $Re=200$ & 500 \\
~ & ~ & $Re=300$ & 500 \\
~ & ~ & $Re=400$ & 500 \\
\hline
DNS-turbulence    & A homogeneous and isotropic turbulence flow   & -  & 2000  \\
\hline
SQG    & Sea surface flow driven by SQG model   & -  & 2000  \\
\hline
Channel    & Channel flow   & -  & 1600  \\
\hline
MHD1024    & Forced MHD turbulence Coarse   & -  & 1000  \\
\hline
Isotropic1024 Coarse    & JHTDB-isotropic1024 Coarse   & -  & 2000  \\
\hline
Isotropic1024 Fine    & JHTDB-isotropic1024 Fine   & -  & 1000  \\
\hline
Mixing    & JHTDB-Mixing   & -  & 1000  \\
\hline
\end{tabular}
\end{center}
\end{table*}

Training the network model requires a large amount of ground truth data sets to optimize the model parameters.
However, it is difficult to obtain accurate velocity fields from real PIV experiments. 
Hence, a synthetic dataset is used for training and evaluation of the network. 
Fig.~\ref{fig2} describes the generation of the synthetic dataset, which consists of the following three main parts.

\textbf{\textit{Flow fields}}. The dataset includes various flow motions generated by computational fluid dynamics (CFD).
Specifically, data sets such as uniform flow (Uniform), backward stepping flow (Back-step) and vortex shedding over a circular cylinder (Cylinder) are provided in \citet{cai2019particle,cai2019}. Corresponding data sets can be downloaded from the website: \url{https://github.com/shengzesnail/PIV_dataset}.
In addition, DNS-turbulence and surface quasi-geostrophic (SQG) sets are provided in \citet{carlier2005second} and \citet{Resseguier2016}, respectively. 
Other turbulence data sets, such as JHTDB-isotropic1024-hd, JHTDB-mhd1024-hd and JHTDB-channel, are provided in Johns Hopkins Turbulence Databases (JHTDB, \url{http://turbulence.pha.jhu.edu/}).
Detailed descriptions of the dataset are shown in Table~\ref{tab1}. 
Each data set contains velocity fields at successive moments, which constitute the velocity field sequence shown in the top row of Fig.~\ref{fig2}.


\textbf{\textit{Particle images}}. The generation of synthetic particle images is shown in Fig.~\ref{fig2}.
The first particle image is constructed by randomly seeding particles whose intensity satisfies a two-dimensional Gaussian function into the image domain.
Next, the underlying flow motions are extracted from the first ground truth velocity field through velocity inquiry. 
Then the motion field is applied to the particles in the first image to obtain the second particle image.
Repeat the above steps to extract the underlying flow motions from the second flow field, and apply them to the particles in the second image to generate the third particle image, and so on.
Eventually, a sequence of particle images can be obtained, as shown in the middle row of Fig.~\ref{fig2}.
The resolution of the generated particle images is 256 $\times$ 256 pixels.
Parameters for defining a particle image are randomly selected in a proper range, as shown in Table~\ref{tab2}.
All particle images are added with the Gaussian noise, which has a sigma uniformly sampled from [0, 0.04].
Starting with different randomly generated particle images, a large number of particle image sequences can be generated based on the velocity field sequences of different flow motions in the dataset. 
Basically we follow the strategy in \citet{cai2019} to generate synthetic particle images. Please refer to this article for more details.

\begin{table}[h]
\begin{center}
\caption{The ranges of parameters for determining a synthetic particle image}\label{tab2}%
\begin{tabular}{ p{3cm} p{2cm} p{2cm}  }
\hline
Parameter & Range  & Unit \\
\hline
Seeding density $\rho$    & 0.05-0.1 & ppp  \\
Particle diameter $d_{p}$    & 1-4 & pixel  \\
Peak intensity $I_{0}$    & 200-255 & grey value  \\
\hline
\end{tabular}
\end{center}
\end{table}

\textbf{\textit{Initial velocity field}}. 
As shown at the bottom of Fig.~\ref{fig2}, the coarse velocity field sequence can be obtained by applying the cross-correlation method to two adjacent particle images in the image sequence.
The commercial software MicroVec 3.6.2 (from MicroVec, Inc.) is used for executing the correlation calculation. 
The window size of the cross-correlation method is 16 $\times$ 16 pixels, and the overlap rate is 50\%, leading to a vector number of the initial velocity field of 32 $\times$ 32. In theory, a coarser initial velocity field can be used, which can further reduce the computational cost of training.

\section{Results and discussions}
\label{sect4}
As introduced in Section~\ref{sect3}, the neural network model is trained using a synthetic PIV dataset. To further demonstrate the performance and practicality of the particle image velocity measurement algorithm based on deep neural network, test results of the proposed CC-FCN with both synthetic and real PIV experimental data are presented in Section~\ref{sect4}.
Due to the various sizes of particle images taken in real PIV experiments, we have developed a set of image segmentation and merging codes following the strategy in \citet{cai2019,cai2019particle}.
In specific, the original input particle images are firstly divided into several parts with the standard size of 256 $\times$ 256 pixels. Then, these parts are processed using CC-FCN and corresponding partial velocity fields are obtained. Finally, the calculation results of each part are orderly combined to get the ultimate output, which is the same size as the original input.

\subsection{Test on spatial accuracy and precision}
\label{sect4.1}

\begin{figure*}
	\begin{center}
	    \subfigure[]{			\includegraphics[height=.35\textwidth]{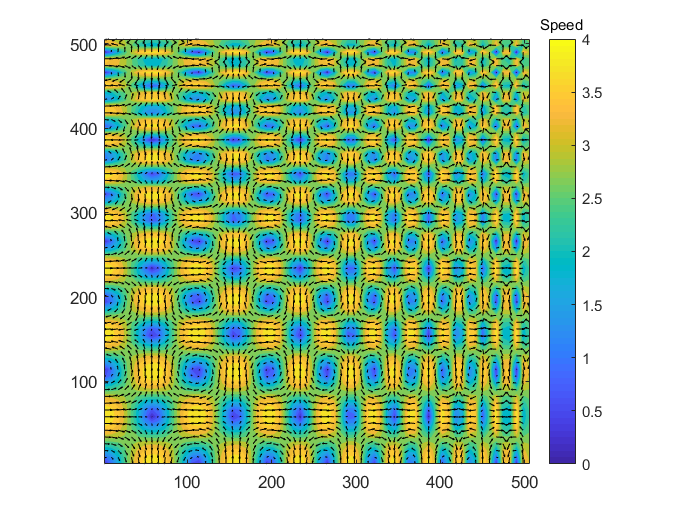}
			\label{fig3a}}
	    \subfigure[]{			\includegraphics[height=.35\textwidth]{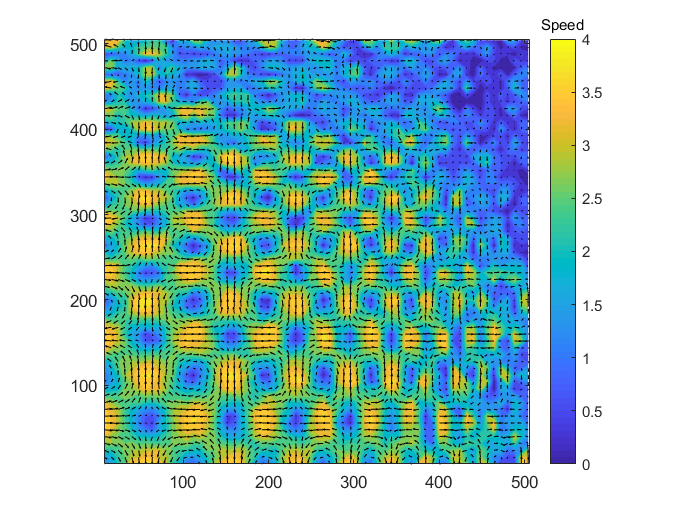}
			\label{fig3b}}\\
		\subfigure[]{			\includegraphics[height=.35\textwidth]{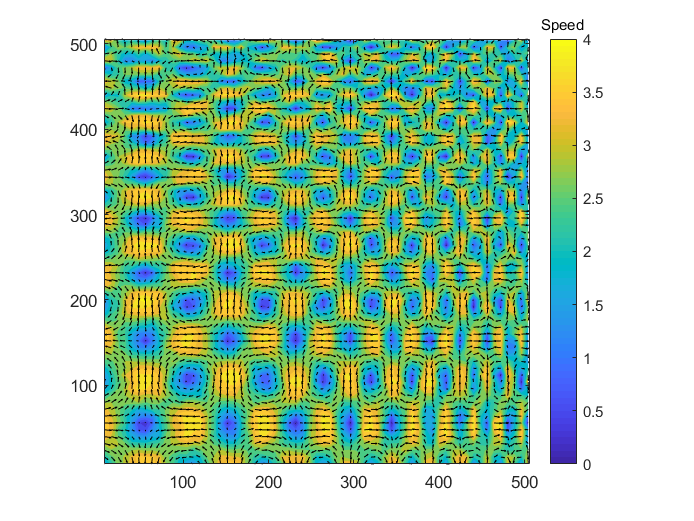}
			\label{fig3c}}
	    \subfigure[]{			\includegraphics[height=.35\textwidth]{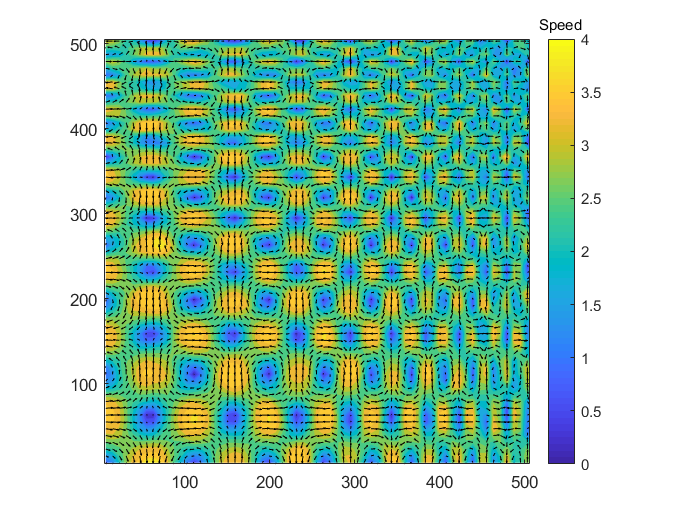}
			\label{fig3d}}
		\caption{Estimated velocity fields from (a) the ground truth, (b) the cross-correlation method, (c) the CC-FCN model (d) the LiteFlowNet-en model. The color map demonstrates the velocity magnitude.}
		\label{fig3}
	\end{center}
\end{figure*}

\begin{figure*}[t]
	\begin{center}
	    \subfigure[]{			\includegraphics[height=.24\textwidth]{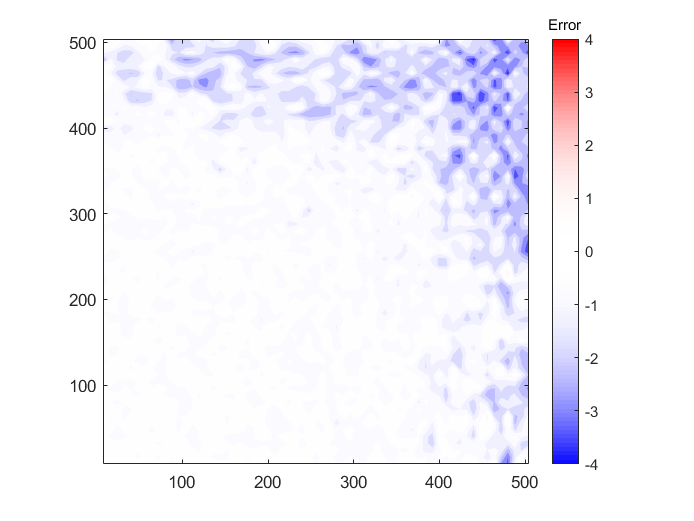}
			\label{fig4a}}
	    \subfigure[]{			\includegraphics[height=.24\textwidth]{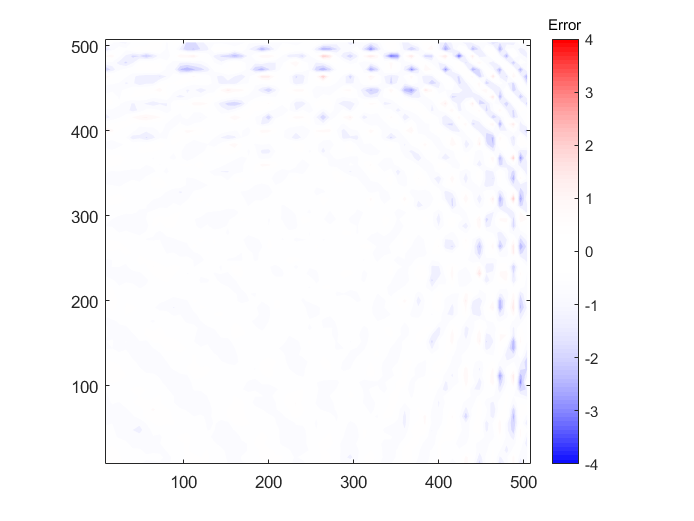}
			\label{fig4b}}
		\subfigure[]{			\includegraphics[height=.24\textwidth]{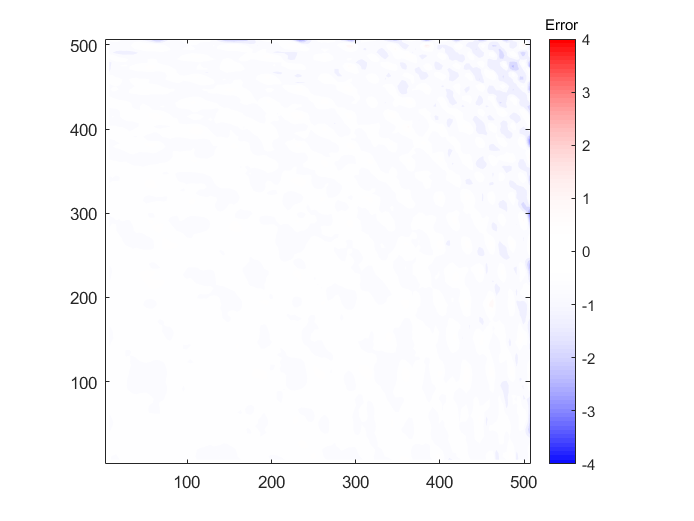}
			\label{fig4c}}
		\caption{Error distributions of (a) the cross-correlation algorithm, (b) the CC-FCN model, (c) the LiteFlowNet-en model.}
		\label{fig4}
	\end{center}
\end{figure*}

\begin{figure*}
	\begin{center}
	    \subfigure[]{			\includegraphics[height=.24\textwidth]{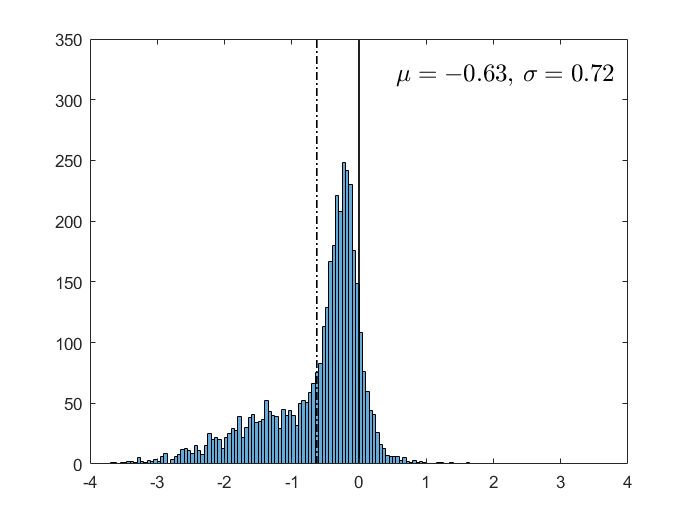}
			\label{fig5a}}
	    \subfigure[]{			\includegraphics[height=.24\textwidth]{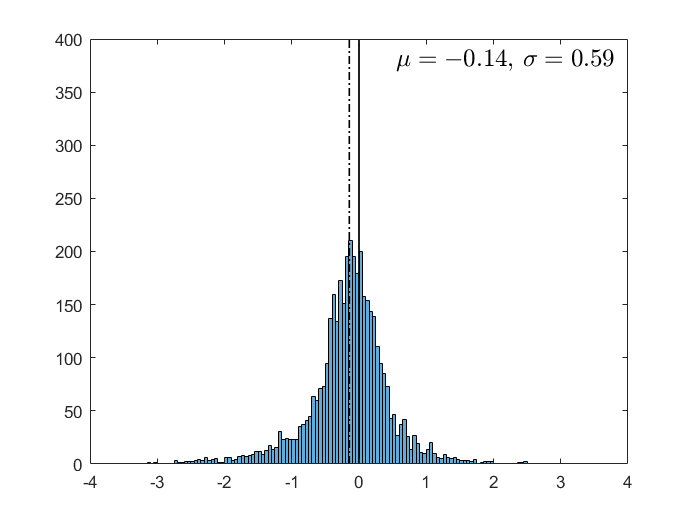}
			\label{fig5b}}
		\subfigure[]{			\includegraphics[height=.24\textwidth]{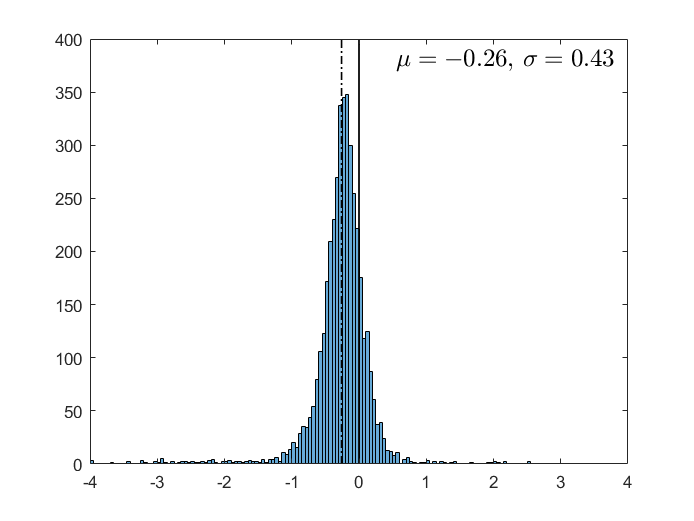}
			\label{fig5c}}
		\caption{Error distribution histograms of (a) the cross-correlation algorithm, (b) the CC-FCN model, (c) the LiteFlowNet-en model.}
		\label{fig5}
	\end{center}
\end{figure*}

Accuracy represents the closeness of the measurement values to the ground truth, and precision exhibits the closeness of the measurement values to each other.
To quantitatively demonstrate the accuracy and precision of the proposed CC-FCN model, we first investigate the model on synthetic particle images. The synthetic particle images are generated from the velocity field as follows:
\begin{equation}
\begin{split}
u(i,j)&=4cos\left[\frac{2\pi i}{256-0.3i}\right]sin\left[\frac{2\pi j}{256-0.3j}\right],\\
v(i,j)&=-4sin\left[\frac{2\pi i}{256-0.3i}\right]cos\left[\frac{2\pi j}{256-0.3j}\right],
\label{eqn8}
\end{split}
\end{equation}
where $i$ and $j$ denote the index of the image pixel along $x$- and $y$-direction, respectively.
These synthetic particle images correspond to vortical flows with varying wavelength, which constitute a standard test case in the PIV community.
For comparison, we also demonstrate the results of the cross-correlation method with WIDIM and LiteFlowNet-en. 
The cross-correlation method uses interrogation window of 16 $\times$ 16 pixels and step size of 8 pixels.
The result of the cross-correlation method is regarded as the baseline of PIV estimation.
The LiteFlowNet-en model is a velocity measurement algorithm based on deep neural network developed by \citet{cai2019particle}, and its result is regarded as the baseline of traditional CNN-based model's estimation.
In the following test cases, estimation results of the cross-correlation method and LiteFlowNet-en are also used as benchmarks for comparison.
The ground truth motion field is shown in Fig.~\ref{fig3}(a), which contains a series of vortices of different sizes. 
Figs.~\ref{fig3}(b)-(d) show the results estimated by the traditional cross-correlation algorithm, the CC-FCN model and the LiteFlowNet-en model, respectively.
The cross-correlation algorithm (Fig.~\ref{fig3}b) is able to capture vortices at different scales, but it gives a relatively poor estimation of speed magnitude in the region of small-scale vortices.
However, the small-scale structures can be estimated very well by CC-FCN and LiteFlowNet-en, as shown in Figs.~\ref{fig3}(c)-(d).
As for the estimation of large-scale vortices, the results obtained by these methods are consistent with the ground truth.
It is worth mentioning that the vector number of the results estimated by CC-FCN and LiteFlowNet-en is 512 $\times$ 512, while the vector number of the flow field estimated by the traditional cross-correlation method is only 63 $\times$ 63.
Although testing on spatial resolution is not the topic of this sub-section, it is not difficult to see that the deep learning based PIV methods show great advantages in spatial resolution as they are single-pixel algorithms. Because it depends on the size of interrogation window, the estimation resolution of the cross-correlation method is inevitably lower than that of single-pixel algorithms.

In order to further compare the calculation quality, we quantify the error of the above methods.
Due to the resolution difference among the results, we first convert the vector number of the flow fields to the same value as 63 $\times$ 63.
Then the error distribution fields are obtained by subtracting the ground truth velocity field from the predicted velocity fields obtained by these three methods. 
The error distributions are shown in Fig.~\ref{fig4}.
It is obvious that the error of the cross-correlation method is quite larger than that of the other two methods for estimating the small-scale vortices. 
The cross-correlation method underestimates the velocity magnitude of small-scale vortices regions.
As shown in Fig.~\ref{fig4}(b), the estimation error is greatly reduced in the same region, which means that CC-FCN can properly extract small-scale flow motions.
Still, more training data related to the flow patterns near the top and right boundaries could be considered.
Fig.~\ref{fig4}(c) indicates that LiteFlowNet-en gives an overall outstanding estimation.

As a quantitative supplement, Fig.~\ref{fig5} shows the histogram of the error of these methods. The mean value and standard deviation of the errors are also denoted in the figure.
For a more intuitive comparison, black dotted and solid lines are used to exhibit the mean error and zero, respectively.
Due to the underestimation of the velocity in the small-scale vortices region, the mean value of the error of the cross-correlation method is the worst away from zero. Additionally, the error of the cross-correlation method has the largest standard deviation, which means that the distribution of error is relatively scattered.
By contrast, CC-FCN can correctly extract flow motions in regions where the cross-correlation estimation is poor, even though the proposed model is based on the result of cross-correlation.
The mean error of CC-FCN is greatly reduced, which is the one with the least deviation from zero. 
It means that CC-FCN is the most accurate approach among the three methods.
The error standard deviation of CC-FCN is also smaller than that of cross-correlation.
Hence, CC-FCN presents an excellent ability to fix defects in the sparse velocity field as input when giving the dense output.
The standard deviation of error of LiteFlowNet-en is the closest to zero, which means that the error distribution of LiteFlowNet-en is less scattered. 
But the corresponding mean error of LiteFlowNet-en is larger than that of CC-FCN, so the result calculated by LiteFlowNet-en is less close to the ground truth.
In summary, CC-FCN has the highest accuracy but its precision cannot outperform the LiteFlowNet-en model.

\begin{figure*}
	\begin{center}
	    \subfigure[]{			\includegraphics[height=.25\textwidth]{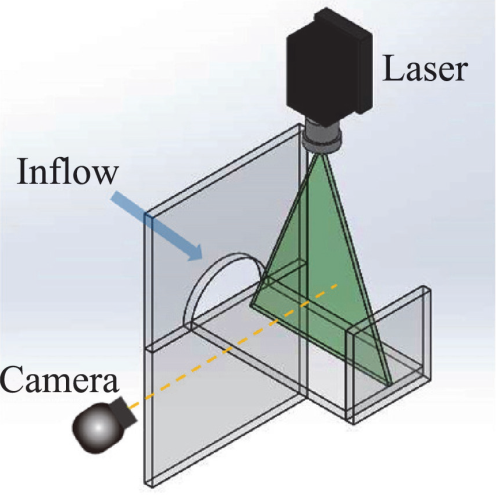}
			\label{fig6a}}
	    \subfigure[]{			\includegraphics[height=.25\textwidth]{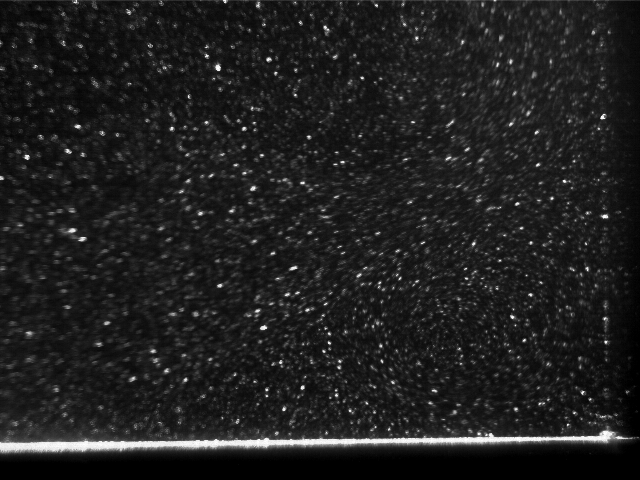}
			\label{fig6b}}
		\caption{(a) Schematic diagram of the jet flow experiment and (b) a particle image sample.}
		\label{fig6}
	\end{center}
\end{figure*}

\begin{figure*}
\begin{center}
  \includegraphics[width=1\textwidth]{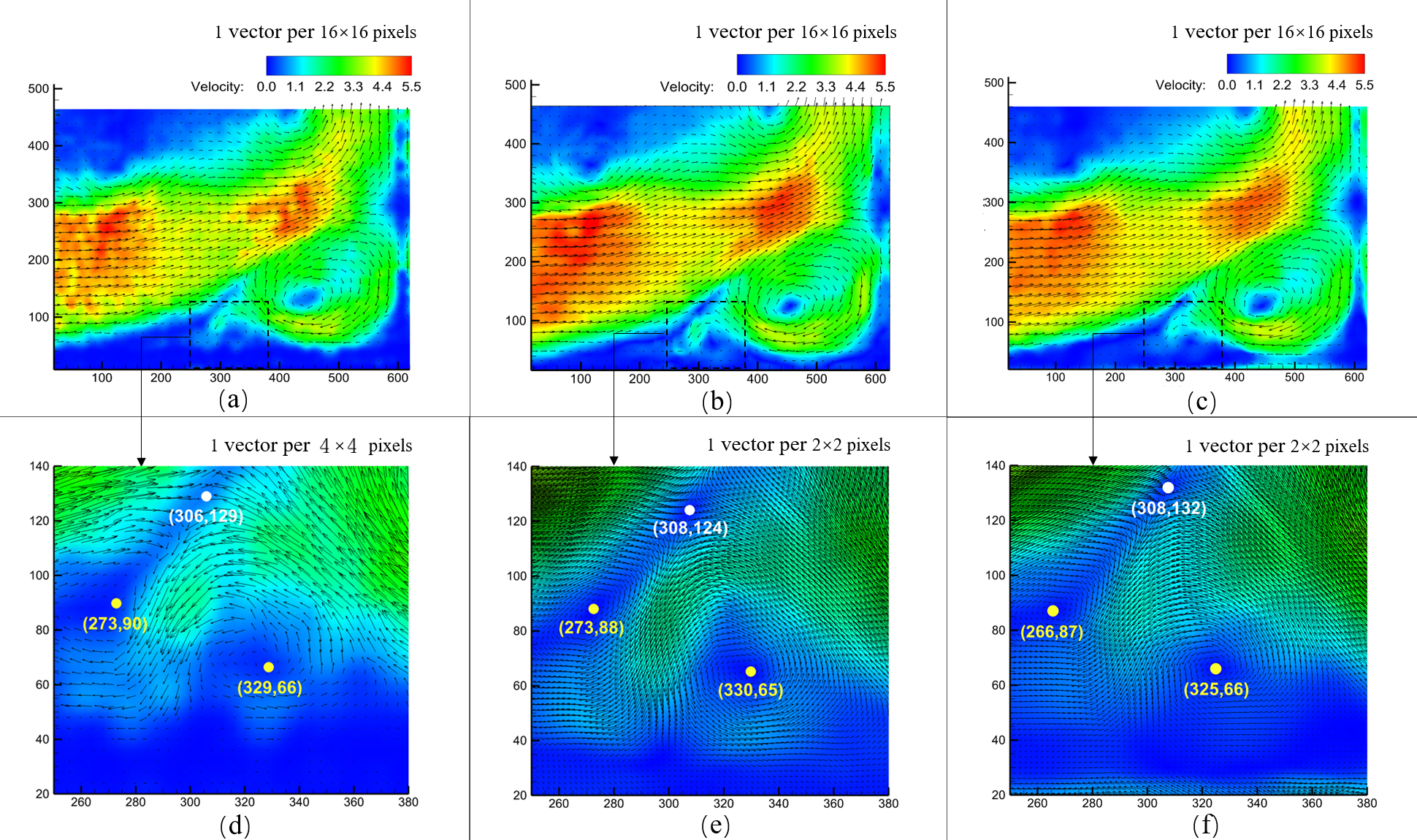}
\caption{Estimated jet flow velocity fields from (a) the cross-correlation method, (b) the proposed CC-FCN model, (c) the LiteFlowNet-en model. Corresponding partial enlargements of the secondary vortex region are shown in (d)-(f), respectively. The color map demonstrates the velocity magnitude. The centers of the two near-wall vortices and the saddle are marked with yellow and white dots, respectively.}
\label{fig7}   
\end{center}
\end{figure*}

\begin{figure*}
	\begin{center}
	    \subfigure[]{			\includegraphics[height=.2\textwidth]{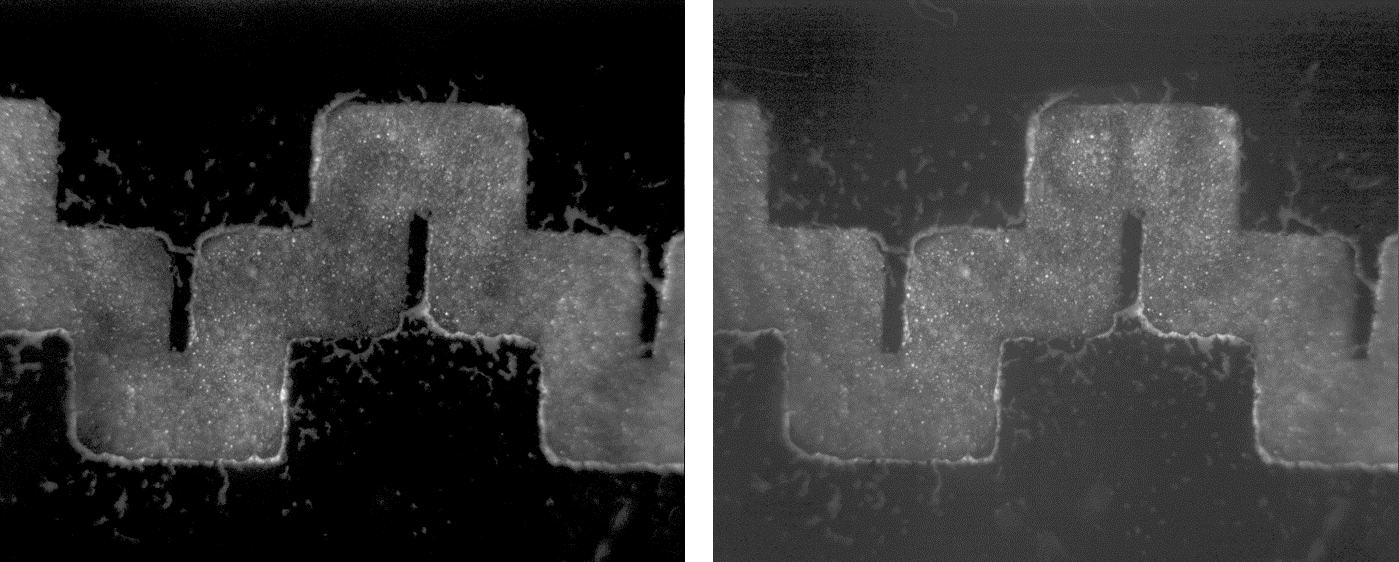}
			\label{fig8a}}
	    \subfigure[]{			\includegraphics[height=.2\textwidth]{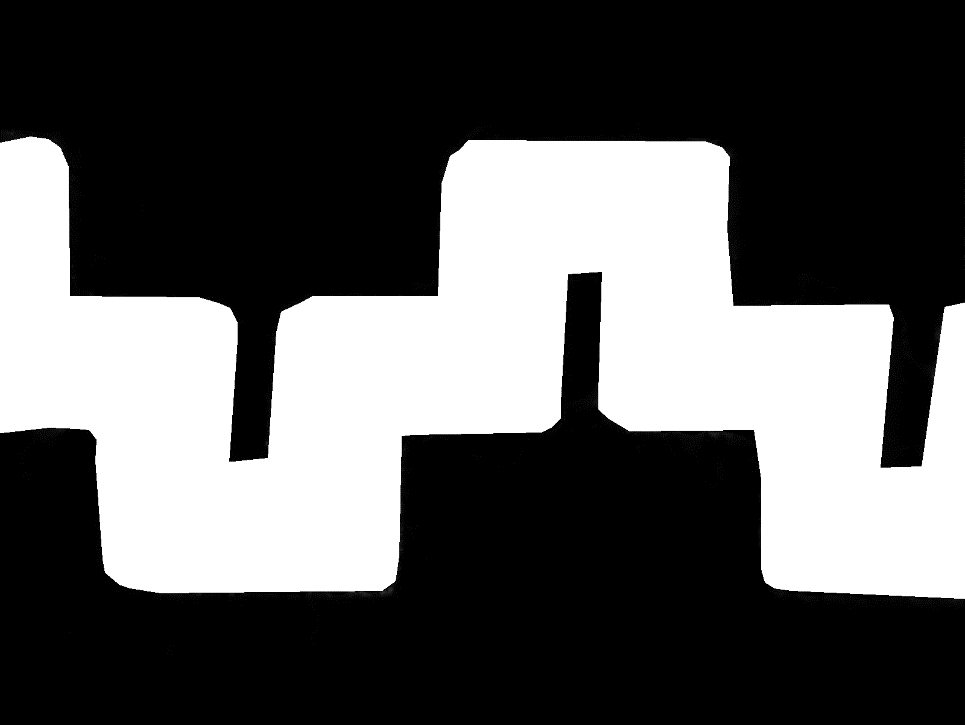}
			\label{fig8b}}
		\caption{(a) Particle images at adjacent times of the agricultural dripper micro-channel PIV experiment and (b) mask of the agricultural dripper micro-channel test. Note that the brightness of the sample images in (a) is increased for a better display. The brightness of the particle images input to the AI-based models is not adjusted.}
		\label{fig8}
	\end{center}
\end{figure*}

\begin{figure*}
\begin{center}
  \includegraphics[width=1\textwidth]{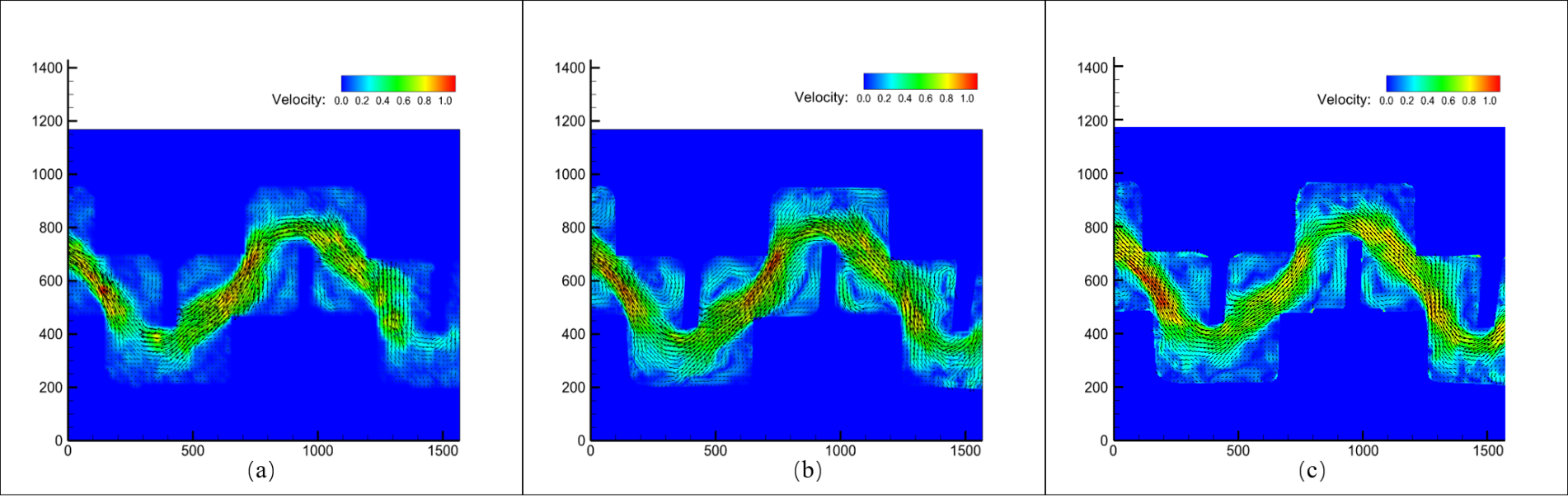}
\caption{Estimated micro-channel flow velocity fields from (a) the cross-correlation method, (b) the proposed CC-FCN model, (c) the LiteFlowNet-en model. The color map demonstrates the velocity magnitude.}
\label{fig10}   
\end{center}
\end{figure*}

\subsection{Test on spatial resolution}
\label{sect4.2}
The jet flow experiment is a common configuration in experimental fluid mechanics, which can form typical flow structures such as jet, velocity gradient change, vortex, separation flow, etc. 
We adopt the jet flow to assess the effective spatial resolution of the CC-FCN model.
Fig.~\ref{fig6} shows the schematic and a particle image sample of the jet flow PIV experiment.
When the incoming flow passes through the upstream semi-circular small hole, a jet is formed. A vertical plate is placed downstream, which is used to induce separation and recirculation flow.
Particles are illuminated by a semiconductor continuous laser (5 W, 532 $\mu$m wavelength), and particle images (640 $\times$ 480 pixels) are recorded by a CCD camera.
The window size and step size of the cross-correlation method are 16 $\times$ 16 pixels and 4 pixels, respectively.

Figs.~\ref{fig7}(a)-(c) present the velocity fields estimated by the cross-correlation method, CC-FCN and LiteFlowNet-en, respectively.
For the sake of comparison, we adjust the resolution of the velocity fields to the same level. Namely, sparse velocity fields with 16 $\times$ 16 pixels for one vector are shown in the above figures.
Thanks to the high-quality particle images obtained in the PIV experiment, all three methods obtain overall good velocity field estimation results.
In order to further compare details of the flow field extracted by these three methods, partial enlargements of a typical near-wall region (marked by the black dotted box) with abundant flow structures are presented in Figs.~\ref{fig7}(d)-(f).
For simplicity of observation, the two partial enlargements of AI-based models are presented with 2 $\times$ 2 pixels for one vector. The velocity field in Fig.~\ref{fig7}(d) is 4 $\times$ 4 pixels for one vector, since  the cross-correlation method can only give sparse estimation.
Nevertheless, as a mature flow motion estimation technology, the cross-correlation method could extract flow structures with relatively high reliability.
Two vortices and one saddle point in the near-wall region are detected by all three methods. Vortex centers and the saddle are marked by yellow and white dots, respectively. 
The coordinate locations are labeled below the dots, through which we can see that the spatial information of typical structures (i.e., two vortex centers and one saddle) captured by CC-FCN is closer to that obtained by cross-correlation rather than by LiteFlowNet-en. 
CC-FCN and the cross-correlation method extract the relatively high velocity around the vortices. Obviously, the former gives more details of the high-speed regions.
By contrast, the flow field extracted by LiteFlowNet-en exhibits smaller high-speed regions around the two vortices, but an abnormal high-speed region close to the wall.
Combining the case shown in Section~\ref{sect4.1}, it can be seen that the deep learning based PIV methods are superior than the traditional cross-correlation method in terms of spatial resolution.
Compared with LiteFlowNet-en, the proposed CC-FCN model can provide the dense velocity field closer to the reliable estimation given by the cross-correlation algorithm.

\subsection{Test on robustness}
\label{sect4.3}

\begin{figure*}
\begin{center}
  \includegraphics[width=0.5\textwidth]{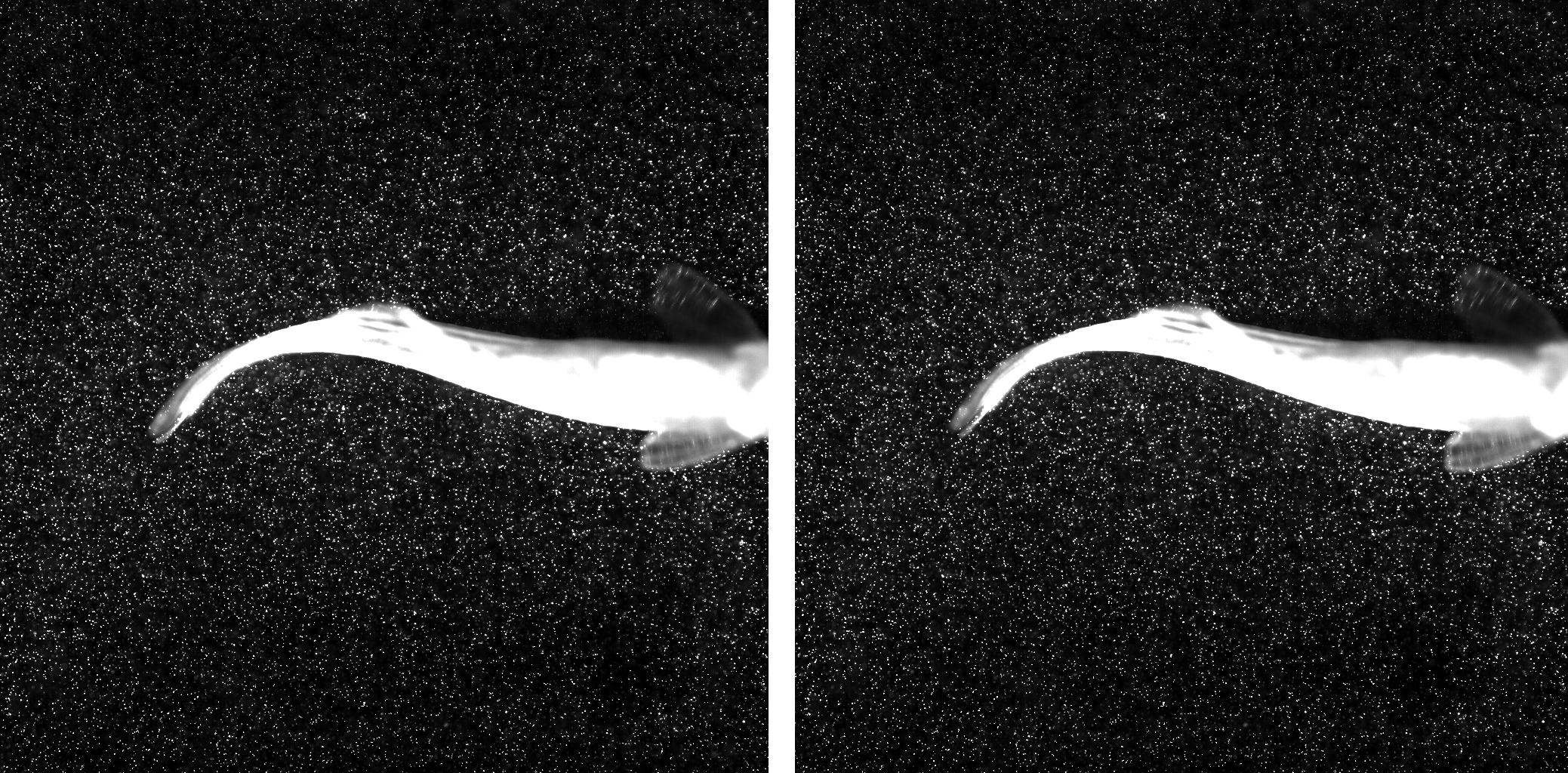}
\caption{Particle images at adjacent times of the zebrafish PIV experiment.}
\label{fig11}   
\end{center}
\end{figure*}

\begin{figure*}
\begin{center}
  \includegraphics[width=0.8\textwidth]{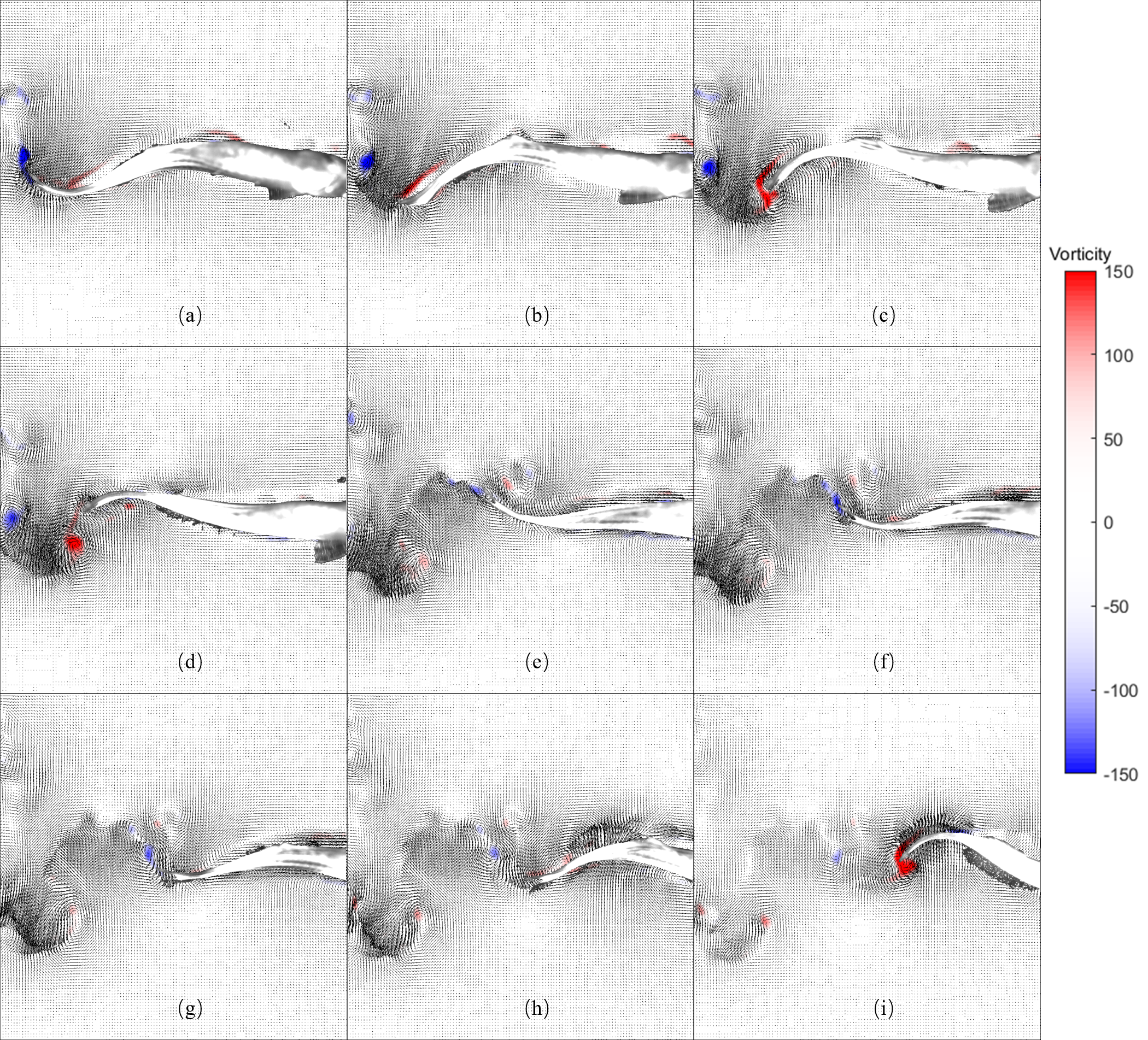}
\caption{Velocity and vorticity fields estimated by the CC-FCN model at different moments. The color map demonstrates the vorticity of the flows.}
\label{fig12}   
\end{center}
\end{figure*}

\begin{figure*}
	\begin{center}
	    \subfigure[]{			\includegraphics[height=.28\textwidth]{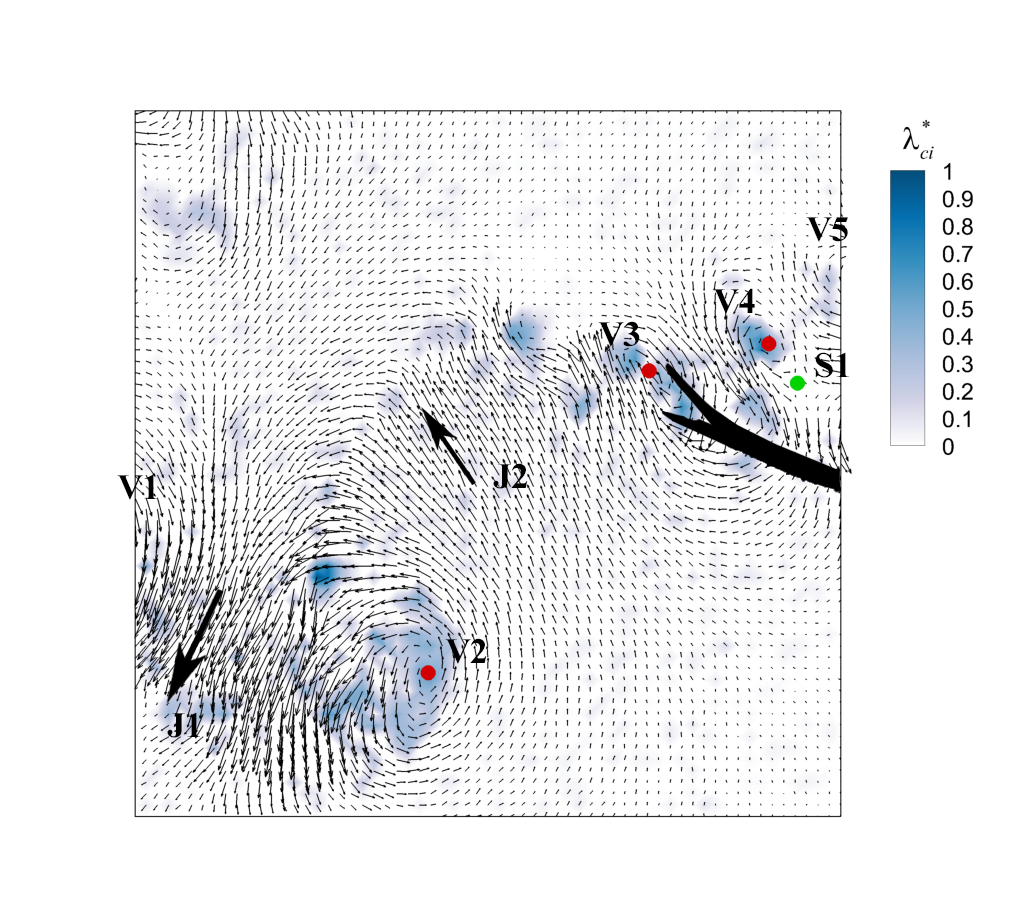}
			\label{fig13a}}
	    \subfigure[]{			\includegraphics[height=.28\textwidth]{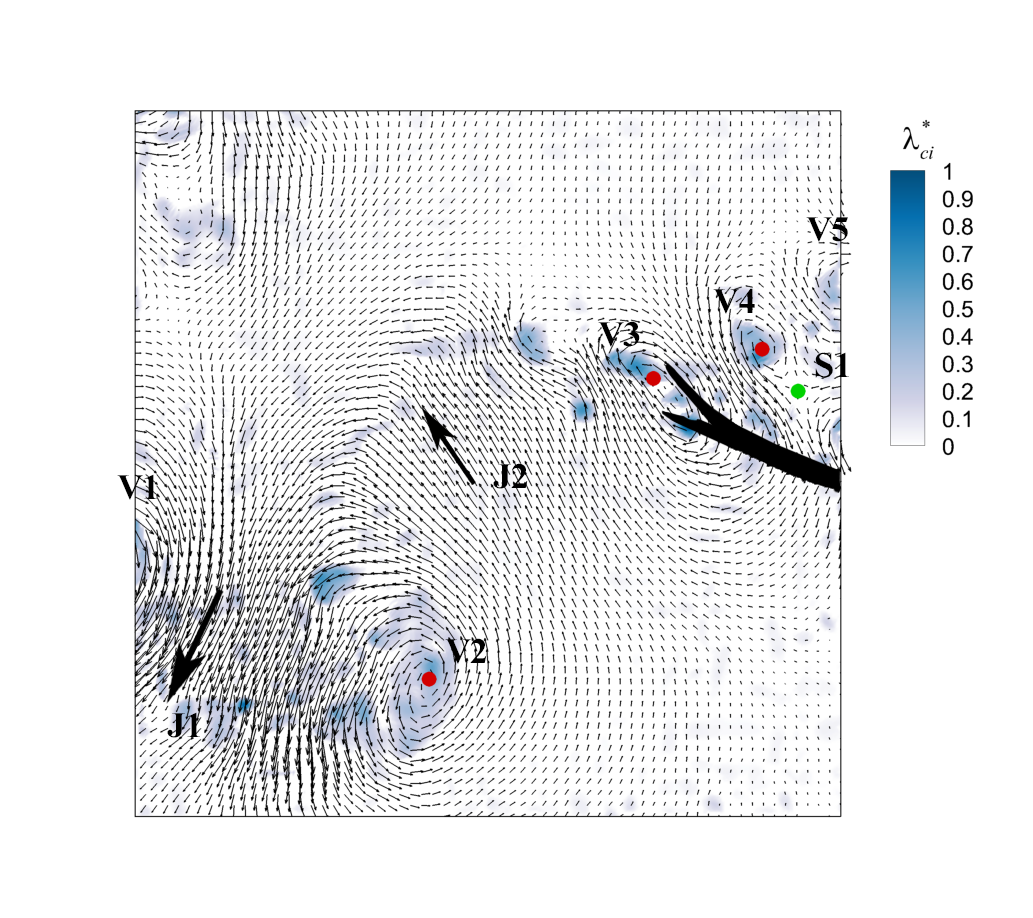}
			\label{fig13b}}
		\subfigure[]{			\includegraphics[height=.28\textwidth]{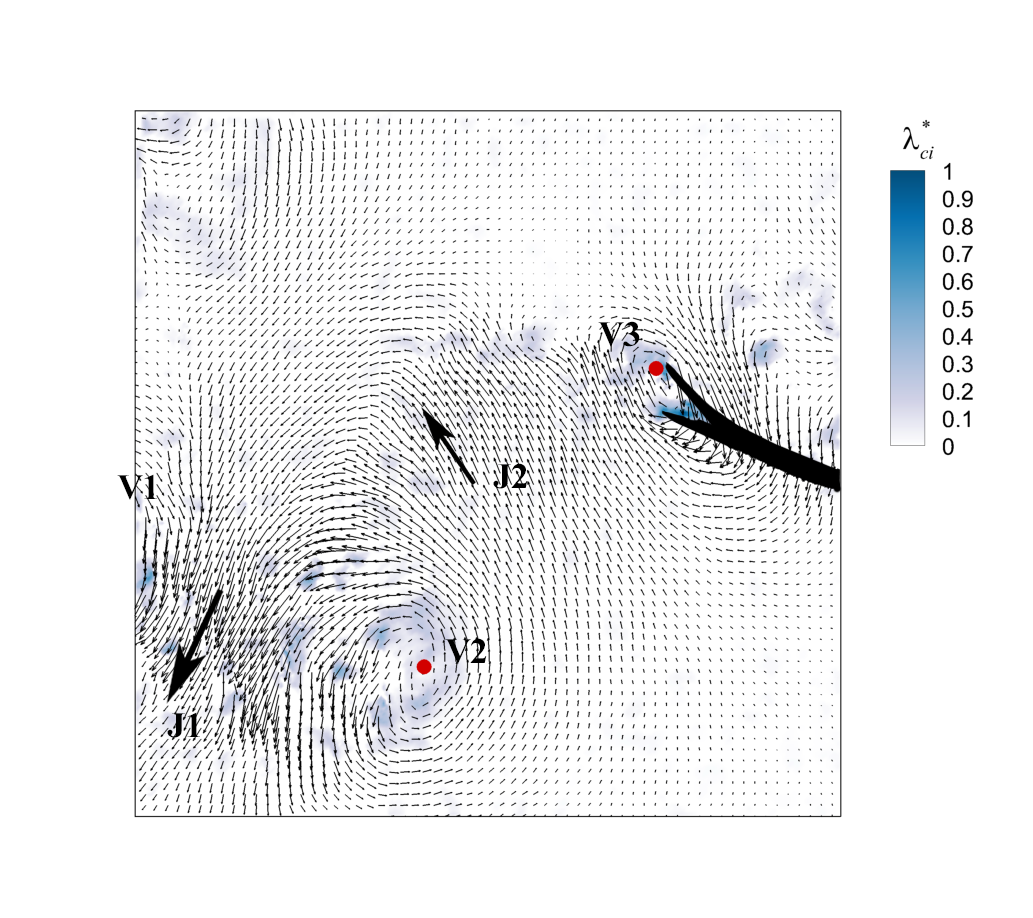}
			\label{fig13c}}
		\caption{Estimated velocity fields at the tail of zebrafish from (a) the cross-correlation method, (b) the CC-FCN model, (c) the LiteFlowNet-en model. The color map demonstrates the velocity magnitude. The color map demonstrates the dimensionless $\lambda_{ci}^{\ast}$. Vortices are denoted from V1 to V5, and their centers are marked with red dots. The saddle near the tail fin (black mask) is denoted as S1 and it is marked with the green dot. Orientations of jets J1 and J2 are indicated by the thick black arrows.}
		\label{fig13}
	\end{center}
\end{figure*}

Robustness is the ability of an algorithm to resist or overcome adverse conditions.
Robustness directly determines the feasibility of the algorithm for practical applications.
In Section~\ref{sect4.1} and Section~\ref{sect4.2}, high-quality particle images are used in the spatial accuracy and resolution test cases.
However, due to the limitation of experimental conditions, the quality of particle images in many PIV experiments may be poor. 
Therefore, it is necessary to investigate robustness of the current proposed model. 
Fig.~\ref{fig8}(a) shows the particle image samples for the robustness test, which is a PIV experiment of agricultural dripper micro-channel. 
Water only flows in the micro-channel, so there are no tracer particles in the area outside the micro-channel.
Particle images (1600 $\times$ 1200 pixels) are captured by a CCD camera equipped with a SM-MICROL-X10 lens.
Illumination is provided by a dual-head
Nd:YAG laser (200 mJ/pulse, 532 nm wavelength), and fluorescent particles about 8 $\mu$m diameter are applied.
The cross-correlation method employs interrogation window of 32 $\times$ 32 pixels and step size of 16 pixels.
A mask (Fig.~\ref{fig8}b) is used to filter the calculation result, so that only the velocity vectors in the white area are displayed. The velocity in the remaining areas is set to zero.

Figs.~\ref{fig10}(a)-(c) show the velocity fields estimated by the cross-correlation method, CC-FCN and LiteFlowNet-en, respectively.
Similar to Section~\ref{sect4.2}, we adjust the resolution of the velocity fields to the same level (16 $\times$ 16 pixels for one vector).
It can be clearly seen from Figs.~\ref{fig10}(a)-(c) that all three methods can resist severe noise and obtain the correct flow motion pattern.
In other words, CC-FCN and LiteFlowNet-en present good robustness when estimating this complicated flow field with a complex boundary.
Small-scale flow motions, such as recirculation flows at the corners, are extracted by these two AI-based models.
Especially, CC-FCN clearly captures structures of those vortical flows with abundant details, including nearby relatively high-speed flows and velocity gradients.
Compared with the cross-correlation result, however, recirculation flows at corners extracted by AI-based models exhibit a higher velocity magnitude.
Meanwhile, the high-speed regions (red/orange contours) extracted by LiteFlowNet-en are slightly different to the corresponding regions detected by the cross-correlation method and CC-FCN.
In specific, high-speed regions in Fig.~\ref{fig10}(c) have a significantly higher velocity magnitude in the left and a lower velocity magnitude in the middle of the micro-channel.
Such distribution may hint an overestimation or underestimation of the velocity magnitude.
It must be mentioned that the brightness of the original input particle images is much lower than that of the sample images shown in Fig.~\ref{fig8}(c).
Hence, the current test also shows that CC-FCN and LiteFlowNet-en are robust against poor illumination.



The above testing cases show that the current proposed CC-FCN model has many advantages.
In order to further evaluate the feasibility of CC-FCN in practical applications, we employ a more complex PIV experiment case to examine the performance of CC-FCN.
The experiment aims to extract the wake flows generated by an adult zebrafish, which freely swims in a water tank filled with fresh water. 
Particle images (1024 $\times$ 1024 pixels) are recorded by a high-speed camera with a 200 mm Nikon macro lens. Particles with a mean diameter of 20 $\mu$m are illuminated by a dual-head high-speed ND:YAG laser (50 mJ/pulse, 527 nm wavelength).
The window size and step size of the cross-correlation method are 16 $\times$ 16 pixels and 8 pixels, respectively. 
Fig.~\ref{fig11} shows a pair of particle images at adjacent times in the particle image sequence. 
The whole particle image sequence consists of 140 particle images.

Fig.~\ref{fig12} exhibits some typical frames of the velocity field sequence calculated by CC-FCN. Velocity vectors are superimposed on the vorticity contours, where red represents positive vorticity and blue represents negative vorticity.
Vorticity accumulates around the tail fin tip when the fish oscillates its tail fin. As a consequence, a vortex is generated and temporally attached to the fin. The vortex gradually grows and finally it is shed into the wake.
As the fish continues to oscillate the fin, vortices are shed in order, and a so-called reverse von K\'{a}rm\'{a}n street \citep{Triantafyllou1993} is formed.
The vortex street and induced jets are clearly detected by CC-FCN.
It is obvious that the CC-FCN model can resist noise and capture flow structures with details for complex practical scenarios, such as the flow field near a zebrafish.

As mentioned earlier, the robustness of the cross-correlation algorithm is very strong. To compare the robustness of cross-correlation, CC-FCN and LiteFlowNet-en in this particular case, we extract two adjacent particle images at a certain moment in the particle image sequence and process them by these three methods, as shown in Fig.~\ref{fig13}.
Dimensionless $\lambda_{ci}^{\ast}$ ($\lambda_{ci}^{\ast}=\lambda_{ci}/\lambda_{ci,max}$) contours are superimposed with velocity vectors. 
The tail fin that sweeps down is represented by the black mask.
Vortices in the wake are denoted from V1 to V5 and their centers are marked with red dots. The saddle point (green dot) near the tail fin is labeled as S1. 
The high-speed flows J1 and J2 induced by the tail fin and the vortex ring composed of V1 and V2 are clearly captured by all three methods, though the structure of J1 and V2 slightly varies from Figs.~\ref{fig13}(a)-(c).
Specifically, the $\lambda_{ci}^{\ast}$ magnitude of V2 and the high-speed region corresponding to J1 extracted by LiteFlowNet-en are obviously smaller than that extracted by the cross-correlation method and CC-FCN. 
The swirling strength of V2 and the momentum of J1 may be underestimated by LiteFlowNet-en.
We believe that in this case, complex flow structures, boundaries with complicated shapes (i.e., the fish tail) and noise in the near-body regions bring challenges to the extraction of velocity fields.
The LiteFlowNet-en model successfully detects the small vortex V3 attached to the tail fin. CC-FCN can extract a more satisfactory deduced velocity field, since it detects the other two small near-body vortices (V4 and V5) and the saddle point S1 under the influence of above-mentioned negative factors.
Again, the locations of vortex centers and saddle captured by CC-FCN are closer to that obtained by cross-correlation rather than by LiteFlowNet-en. 
The test of the zebrafish experiment again ensures the good robustness of CC-FCN and its feasibility to complex practical applications. 




\section{Conclusion}
\label{sect5}
In this paper, we propose a new neural network structure, which is named as CC-FCN.
Considering that the cross-correlation method is fairly robust and the deep learning method can obtain super-resolution estimation, CC-FCN is designed to achieve a synergetic combination of the cross-correlation method and deep learning. Their advantages (e.g., robustness, super-resolution and high-accuracy) are inherited by the CC-FCN model.
This neural network structure has two types of inputs, one is the particle images and the other is the low-resolution initial velocity field. 
The output of the neural network is a dense velocity field with single-pixel-resolution. 
As the supervised learning strategy is considered, a synthetic dataset including various ground-truth fluid motions is generated to train the parameters of the network. 
Finally, CC-FCN is tested with the real PIV experimental data. The velocity fields estimated by CC-FCN are compared with the results obtained by the cross-correlation method and LiteFlowNet-en, which is a state-of-the-art deep learning based velocity estimation method. 
The proposed CC-FCN model gives flow estimation with the highest accuracy in the synthetic multiple vortices test. And it also exhibits good robustness in practical experimental tests, which involve complicated flow structures with complex inner/outer boundaries commonly seen in PIV experiments.
Slightly differences exist in the estimations of these three methods in experimental tests, but the estimations of CC-FCN and the cross-correlation algorithm (namely, a reliable reference) have greater closeness.
The proposed neural network can properly fuse the two kinds of input information and successfully achieve the competitive super-resolution estimation.


\section*{Acknowledgment}
This work was supported by the National Key Research and Development Program of China (Grant No. 2020YFA040070), the National Natural Science Foundation of China (Grant No. 12072320) and the State Key Program of National Natural Science Foundation of China (Grant No. 91852204).
The authors would like to thank Dr. Shengze Cai for his help in data analysis with the LiteFlowNet-en model.

\section*{Data Availability Statement}

The data that support the findings of this study are available from the corresponding author upon reasonable request.

\bibliography{AIbib}

\begin{thebibliography}{45}%
\makeatletter
\providecommand \@ifxundefined [1]{%
 \@ifx{#1\undefined}
}%
\providecommand \@ifnum [1]{%
 \ifnum #1\expandafter \@firstoftwo
 \else \expandafter \@secondoftwo
 \fi
}%
\providecommand \@ifx [1]{%
 \ifx #1\expandafter \@firstoftwo
 \else \expandafter \@secondoftwo
 \fi
}%
\providecommand \natexlab [1]{#1}%
\providecommand \enquote  [1]{``#1''}%
\providecommand \bibnamefont  [1]{#1}%
\providecommand \bibfnamefont [1]{#1}%
\providecommand \citenamefont [1]{#1}%
\providecommand \href@noop [0]{\@secondoftwo}%
\providecommand \href [0]{\begingroup \@sanitize@url \@href}%
\providecommand \@href[1]{\@@startlink{#1}\@@href}%
\providecommand \@@href[1]{\endgroup#1\@@endlink}%
\providecommand \@sanitize@url [0]{\catcode `\\12\catcode `\$12\catcode
  `\&12\catcode `\#12\catcode `\^12\catcode `\_12\catcode `\%12\relax}%
\providecommand \@@startlink[1]{}%
\providecommand \@@endlink[0]{}%
\providecommand \url  [0]{\begingroup\@sanitize@url \@url }%
\providecommand \@url [1]{\endgroup\@href {#1}{\urlprefix }}%
\providecommand \urlprefix  [0]{URL }%
\providecommand \Eprint [0]{\href }%
\providecommand \doibase [0]{http://dx.doi.org/}%
\providecommand \selectlanguage [0]{\@gobble}%
\providecommand \bibinfo  [0]{\@secondoftwo}%
\providecommand \bibfield  [0]{\@secondoftwo}%
\providecommand \translation [1]{[#1]}%
\providecommand \BibitemOpen [0]{}%
\providecommand \bibitemStop [0]{}%
\providecommand \bibitemNoStop [0]{.\EOS\space}%
\providecommand \EOS [0]{\spacefactor3000\relax}%
\providecommand \BibitemShut  [1]{\csname bibitem#1\endcsname}%
\let\auto@bib@innerbib\@empty
\bibitem [{\citenamefont {Adrian}(1991)}]{Adrian1991}%
  \BibitemOpen
  \bibfield  {author} {\bibinfo {author} {\bibfnamefont {R.}~\bibnamefont
  {Adrian}},\ }\bibfield  {title} {\enquote {\bibinfo {title} {Particle-imaging
  techniques for experimental fluid mechanics},}\ }\href {\doibase
  10.1146/annurev.fl.23.010191.001401} {\bibfield  {journal} {\bibinfo
  {journal} {Annual Review of Fluid Mechanics}\ }\textbf {\bibinfo {volume}
  {23}},\ \bibinfo {pages} {261--304} (\bibinfo {year} {1991})}\BibitemShut
  {NoStop}%
\bibitem [{\citenamefont {Li}\ \emph {et~al.}(2008)\citenamefont {Li},
  \citenamefont {du}, \citenamefont {Zhang},\ and\ \citenamefont
  {Wang}}]{LiAI2008}%
  \BibitemOpen
  \bibfield  {author} {\bibinfo {author} {\bibfnamefont {M.}~\bibnamefont
  {Li}}, \bibinfo {author} {\bibfnamefont {H.}~\bibnamefont {du}}, \bibinfo
  {author} {\bibfnamefont {Q.}~\bibnamefont {Zhang}}, \ and\ \bibinfo {author}
  {\bibfnamefont {J.}~\bibnamefont {Wang}},\ }\bibfield  {title} {\enquote
  {\bibinfo {title} {Improved particle image velocimetry through cell
  segmentation and competitive survival},}\ }\href {\doibase
  10.1109/TIM.2007.915443} {\bibfield  {journal} {\bibinfo  {journal}
  {Instrumentation and Measurement, IEEE Transactions on}\ }\textbf {\bibinfo
  {volume} {57}},\ \bibinfo {pages} {1221 -- 1229} (\bibinfo {year}
  {2008})}\BibitemShut {NoStop}%
\bibitem [{\citenamefont {Meng}, \citenamefont {Li},\ and\ \citenamefont
  {Du}(2009)}]{Juan2009}%
  \BibitemOpen
  \bibfield  {author} {\bibinfo {author} {\bibfnamefont {J.}~\bibnamefont
  {Meng}}, \bibinfo {author} {\bibfnamefont {M.}~\bibnamefont {Li}}, \ and\
  \bibinfo {author} {\bibfnamefont {H.}~\bibnamefont {Du}},\ }\bibfield
  {title} {\enquote {\bibinfo {title} {Particle selection strategy for
  three-dimensional particle image velocimetry},}\ }in\ \href {\doibase
  10.1109/CISP.2009.5304504} {\emph {\bibinfo {booktitle} {2009 2nd
  International Congress on Image and Signal Processing}}}\ (\bibinfo
  {organization} {IEEE},\ \bibinfo {year} {2009})\ pp.\ \bibinfo {pages}
  {1--5}\BibitemShut {NoStop}%
\bibitem [{\citenamefont {Taylor}\ \emph {et~al.}(2011)\citenamefont {Taylor},
  \citenamefont {Gurka}, \citenamefont {Kopp},\ and\ \citenamefont
  {Liberzon}}]{Taylor2011}%
  \BibitemOpen
  \bibfield  {author} {\bibinfo {author} {\bibfnamefont {Z.}~\bibnamefont
  {Taylor}}, \bibinfo {author} {\bibfnamefont {R.}~\bibnamefont {Gurka}},
  \bibinfo {author} {\bibfnamefont {G.}~\bibnamefont {Kopp}}, \ and\ \bibinfo
  {author} {\bibfnamefont {A.}~\bibnamefont {Liberzon}},\ }\bibfield  {title}
  {\enquote {\bibinfo {title} {Long-duration time-resolved {PIV} to study
  unsteady aerodynamics},}\ }\href {\doibase 10.1109/TIM.2010.2047149}
  {\bibfield  {journal} {\bibinfo  {journal} {Instrumentation and Measurement,
  IEEE Transactions on}\ }\textbf {\bibinfo {volume} {59}},\ \bibinfo {pages}
  {3262 -- 3269} (\bibinfo {year} {2011})}\BibitemShut {NoStop}%
\bibitem [{\citenamefont {Triep}, \citenamefont {Br{\"u}cker},\ and\
  \citenamefont {Sie{\ss}}(2006)}]{triep2006}%
  \BibitemOpen
  \bibfield  {author} {\bibinfo {author} {\bibfnamefont {M.}~\bibnamefont
  {Triep}}, \bibinfo {author} {\bibfnamefont {C.}~\bibnamefont {Br{\"u}cker}},
  \ and\ \bibinfo {author} {\bibfnamefont {T.}~\bibnamefont {Sie{\ss}}},\
  }\bibfield  {title} {\enquote {\bibinfo {title} {{DPIV}-measurements of the
  flow field in a micro-axial blood pump},}\ }in\ \href@noop {} {\emph
  {\bibinfo {booktitle} {13th International Symposium on Application of Laser
  Techniques to Fluid Mechanics}}}\ (\bibinfo {organization} {Citeseer},\
  \bibinfo {year} {2006})\BibitemShut {NoStop}%
\bibitem [{\citenamefont {Triep}\ \emph {et~al.}(2008)\citenamefont {Triep},
  \citenamefont {Br{\"u}cker}, \citenamefont {Kerkhoffs}, \citenamefont
  {Schumacher},\ and\ \citenamefont {Marseille}}]{triep2008}%
  \BibitemOpen
  \bibfield  {author} {\bibinfo {author} {\bibfnamefont {M.}~\bibnamefont
  {Triep}}, \bibinfo {author} {\bibfnamefont {C.}~\bibnamefont {Br{\"u}cker}},
  \bibinfo {author} {\bibfnamefont {W.}~\bibnamefont {Kerkhoffs}}, \bibinfo
  {author} {\bibfnamefont {O.}~\bibnamefont {Schumacher}}, \ and\ \bibinfo
  {author} {\bibfnamefont {O.}~\bibnamefont {Marseille}},\ }\bibfield  {title}
  {\enquote {\bibinfo {title} {Investigation of the washout effect in a
  magnetically driven axial blood pump},}\ }\href {\doibase
  10.1111/j.1525-1594.2008.00630.x} {\bibfield  {journal} {\bibinfo  {journal}
  {Artificial organs}\ }\textbf {\bibinfo {volume} {32}},\ \bibinfo {pages}
  {778--784} (\bibinfo {year} {2008})}\BibitemShut {NoStop}%
\bibitem [{\citenamefont {Tytell}\ and\ \citenamefont
  {Lauder}(2008)}]{Tytell2008}%
  \BibitemOpen
  \bibfield  {author} {\bibinfo {author} {\bibfnamefont {E.}~\bibnamefont
  {Tytell}}\ and\ \bibinfo {author} {\bibfnamefont {G.}~\bibnamefont
  {Lauder}},\ }\bibfield  {title} {\enquote {\bibinfo {title} {Hydrodynamics of
  the escape response in bluegill sunfish, {Lepomis} macrochirus},}\ }\href
  {\doibase 10.1242/jeb.020917} {\bibfield  {journal} {\bibinfo  {journal} {The
  Journal of experimental biology}\ }\textbf {\bibinfo {volume} {211}},\
  \bibinfo {pages} {3359--3369} (\bibinfo {year} {2008})}\BibitemShut {NoStop}%
\bibitem [{\citenamefont {Ting}\ and\ \citenamefont {Yang}(2009)}]{Ting2009}%
  \BibitemOpen
  \bibfield  {author} {\bibinfo {author} {\bibfnamefont {S.-C.}\ \bibnamefont
  {Ting}}\ and\ \bibinfo {author} {\bibfnamefont {J.-T.}\ \bibnamefont
  {Yang}},\ }\bibfield  {title} {\enquote {\bibinfo {title} {Extracting
  energetically dominant flow features in a complicated fish wake using
  singular-value decomposition},}\ }\href {\doibase 10.1063/1.3122802}
  {\bibfield  {journal} {\bibinfo  {journal} {Physics of Fluids - PHYS FLUIDS}\
  }\textbf {\bibinfo {volume} {21}} (\bibinfo {year} {2009}),\
  10.1063/1.3122802}\BibitemShut {NoStop}%
\bibitem [{\citenamefont {Flammang}\ \emph {et~al.}(2011)\citenamefont
  {Flammang}, \citenamefont {Lauder}, \citenamefont {Troolin},\ and\
  \citenamefont {Strand}}]{Flammang2011}%
  \BibitemOpen
  \bibfield  {author} {\bibinfo {author} {\bibfnamefont {B.}~\bibnamefont
  {Flammang}}, \bibinfo {author} {\bibfnamefont {G.}~\bibnamefont {Lauder}},
  \bibinfo {author} {\bibfnamefont {D.}~\bibnamefont {Troolin}}, \ and\
  \bibinfo {author} {\bibfnamefont {T.}~\bibnamefont {Strand}},\ }\bibfield
  {title} {\enquote {\bibinfo {title} {Volumetric imaging of fish
  locomotion},}\ }\href {\doibase 10.1098/rsbl.2011.0282} {\bibfield  {journal}
  {\bibinfo  {journal} {Biology letters}\ }\textbf {\bibinfo {volume} {7}},\
  \bibinfo {pages} {695--8} (\bibinfo {year} {2011})}\BibitemShut {NoStop}%
\bibitem [{\citenamefont {Shen}, \citenamefont {Tan},\ and\ \citenamefont
  {Lai}(2012)}]{Shen2012}%
  \BibitemOpen
  \bibfield  {author} {\bibinfo {author} {\bibfnamefont {G.}~\bibnamefont
  {Shen}}, \bibinfo {author} {\bibfnamefont {G.}~\bibnamefont {Tan}}, \ and\
  \bibinfo {author} {\bibfnamefont {G.}~\bibnamefont {Lai}},\ }\bibfield
  {title} {\enquote {\bibinfo {title} {Investigation on {3Dt} wake flow
  structures of swimming bionic fish},}\ }\href {\doibase
  10.1007/s10409-012-0108-5} {\bibfield  {journal} {\bibinfo  {journal} {Acta
  Mechanica Sinica}\ }\textbf {\bibinfo {volume} {28}} (\bibinfo {year}
  {2012}),\ 10.1007/s10409-012-0108-5}\BibitemShut {NoStop}%
\bibitem [{\citenamefont {Wang}\ \emph {et~al.}(2019)\citenamefont {Wang},
  \citenamefont {Gao}, \citenamefont {Wang}, \citenamefont {Wang},\ and\
  \citenamefont {Pan}}]{wang2019experimental}%
  \BibitemOpen
  \bibfield  {author} {\bibinfo {author} {\bibfnamefont {C.}~\bibnamefont
  {Wang}}, \bibinfo {author} {\bibfnamefont {Q.}~\bibnamefont {Gao}}, \bibinfo
  {author} {\bibfnamefont {J.}~\bibnamefont {Wang}}, \bibinfo {author}
  {\bibfnamefont {B.}~\bibnamefont {Wang}}, \ and\ \bibinfo {author}
  {\bibfnamefont {C.}~\bibnamefont {Pan}},\ }\bibfield  {title} {\enquote
  {\bibinfo {title} {Experimental study on dominant vortex structures in
  near-wall region of turbulent boundary layer based on tomographic particle
  image velocimetry},}\ }\href {\doibase 10.1017/jfm.2019.412} {\bibfield
  {journal} {\bibinfo  {journal} {Journal of Fluid Mechanics}\ }\textbf
  {\bibinfo {volume} {874}},\ \bibinfo {pages} {426--454} (\bibinfo {year}
  {2019})}\BibitemShut {NoStop}%
\bibitem [{\citenamefont {Wang}\ \emph {et~al.}(2021)\citenamefont {Wang},
  \citenamefont {Wang}, \citenamefont {Tian},\ and\ \citenamefont
  {Jiang}}]{Wang2021}%
  \BibitemOpen
  \bibfield  {author} {\bibinfo {author} {\bibfnamefont {X.}~\bibnamefont
  {Wang}}, \bibinfo {author} {\bibfnamefont {Y.}~\bibnamefont {Wang}}, \bibinfo
  {author} {\bibfnamefont {H.}~\bibnamefont {Tian}}, \ and\ \bibinfo {author}
  {\bibfnamefont {N.}~\bibnamefont {Jiang}},\ }\bibfield  {title} {\enquote
  {\bibinfo {title} {Effects of the slip wall on the drag and coherent
  structures of turbulent boundary layer},}\ }\href {\doibase
  10.1007/s10409-021-01092-0} {\bibfield  {journal} {\bibinfo  {journal} {Acta
  Mechanica Sinica}\ } (\bibinfo {year} {2021}),\
  10.1007/s10409-021-01092-0}\BibitemShut {NoStop}%
\bibitem [{\citenamefont {Cai}\ \emph {et~al.}(2019{\natexlab{a}})\citenamefont
  {Cai}, \citenamefont {Zhou}, \citenamefont {Xu},\ and\ \citenamefont
  {Gao}}]{cai2019}%
  \BibitemOpen
  \bibfield  {author} {\bibinfo {author} {\bibfnamefont {S.}~\bibnamefont
  {Cai}}, \bibinfo {author} {\bibfnamefont {S.}~\bibnamefont {Zhou}}, \bibinfo
  {author} {\bibfnamefont {C.}~\bibnamefont {Xu}}, \ and\ \bibinfo {author}
  {\bibfnamefont {Q.}~\bibnamefont {Gao}},\ }\bibfield  {title} {\enquote
  {\bibinfo {title} {Dense motion estimation of particle images via a
  convolutional neural network},}\ }\href {\doibase 10.1007/s00348-019-2717-2}
  {\bibfield  {journal} {\bibinfo  {journal} {Experiments in Fluids}\ }\textbf
  {\bibinfo {volume} {60}},\ \bibinfo {pages} {1--16} (\bibinfo {year}
  {2019}{\natexlab{a}})}\BibitemShut {NoStop}%
\bibitem [{\citenamefont {Raffel}\ \emph {et~al.}(2007)\citenamefont {Raffel},
  \citenamefont {Willert}, \citenamefont {Wereley},\ and\ \citenamefont
  {Kompenhans}}]{Raffel2007}%
  \BibitemOpen
  \bibfield  {author} {\bibinfo {author} {\bibfnamefont {M.}~\bibnamefont
  {Raffel}}, \bibinfo {author} {\bibfnamefont {C.}~\bibnamefont {Willert}},
  \bibinfo {author} {\bibfnamefont {S.}~\bibnamefont {Wereley}}, \ and\
  \bibinfo {author} {\bibfnamefont {J.}~\bibnamefont {Kompenhans}},\ }\href
  {\doibase 10.1007/978-3-540-72308-0} {\emph {\bibinfo {title} {Particle image
  velocimetry: a practical guide}}}\ (\bibinfo  {publisher} {Springer},\
  \bibinfo {year} {2007})\BibitemShut {NoStop}%
\bibitem [{\citenamefont {Adrian}\ and\ \citenamefont
  {Westerweel}(2011)}]{Adrian2011}%
  \BibitemOpen
  \bibfield  {author} {\bibinfo {author} {\bibfnamefont {R.}~\bibnamefont
  {Adrian}}\ and\ \bibinfo {author} {\bibfnamefont {J.}~\bibnamefont
  {Westerweel}},\ }\href@noop {} {\emph {\bibinfo {title} {Particle image
  velocimetry}}}\ (\bibinfo  {publisher} {Cambridge university press},\
  \bibinfo {year} {2011})\BibitemShut {NoStop}%
\bibitem [{\citenamefont {Pratt}\ and\ \citenamefont {Kopp}(2013)}]{Pratt2013}%
  \BibitemOpen
  \bibfield  {author} {\bibinfo {author} {\bibfnamefont {R.}~\bibnamefont
  {Pratt}}\ and\ \bibinfo {author} {\bibfnamefont {G.}~\bibnamefont {Kopp}},\
  }\bibfield  {title} {\enquote {\bibinfo {title} {Velocity measurements around
  low-profile, tilted, solar arrays mounted on large flat-roofs, for wall
  normal wind directions},}\ }\href {\doibase 10.1016/j.jweia.2013.09.001}
  {\bibfield  {journal} {\bibinfo  {journal} {Journal of Wind Engineering and
  Industrial Aerodynamics}\ }\textbf {\bibinfo {volume} {123}} (\bibinfo {year}
  {2013}),\ 10.1016/j.jweia.2013.09.001}\BibitemShut {NoStop}%
\bibitem [{\citenamefont {Wang}\ \emph {et~al.}(2017)\citenamefont {Wang},
  \citenamefont {Gao}, \citenamefont {Pan}, \citenamefont {Feng},\ and\
  \citenamefont {Wang}}]{IPTA}%
  \BibitemOpen
  \bibfield  {author} {\bibinfo {author} {\bibfnamefont {Z.}~\bibnamefont
  {Wang}}, \bibinfo {author} {\bibfnamefont {Q.}~\bibnamefont {Gao}}, \bibinfo
  {author} {\bibfnamefont {C.}~\bibnamefont {Pan}}, \bibinfo {author}
  {\bibfnamefont {L.}~\bibnamefont {Feng}}, \ and\ \bibinfo {author}
  {\bibfnamefont {J.}~\bibnamefont {Wang}},\ }\bibfield  {title} {\enquote
  {\bibinfo {title} {Imaginary particle tracking accelerometry based on
  time-resolved velocity fields},}\ }\href {\doibase 10.1007/s00348-017-2394-y}
  {\bibfield  {journal} {\bibinfo  {journal} {Experiments in Fluids}\ }\textbf
  {\bibinfo {volume} {58}} (\bibinfo {year} {2017}),\
  10.1007/s00348-017-2394-y}\BibitemShut {NoStop}%
\bibitem [{\citenamefont {Wang}\ \emph {et~al.}(2016)\citenamefont {Wang},
  \citenamefont {Gao}, \citenamefont {Wang}, \citenamefont {Runjie},\ and\
  \citenamefont {Wang}}]{irro-pre}%
  \BibitemOpen
  \bibfield  {author} {\bibinfo {author} {\bibfnamefont {Z.}~\bibnamefont
  {Wang}}, \bibinfo {author} {\bibfnamefont {Q.}~\bibnamefont {Gao}}, \bibinfo
  {author} {\bibfnamefont {C.}~\bibnamefont {Wang}}, \bibinfo {author}
  {\bibfnamefont {W.}~\bibnamefont {Runjie}}, \ and\ \bibinfo {author}
  {\bibfnamefont {J.}~\bibnamefont {Wang}},\ }\bibfield  {title} {\enquote
  {\bibinfo {title} {An irrotation correction on pressure gradient and
  orthogonal-path integration for {PIV}-based pressure reconstruction},}\
  }\href {\doibase 10.1007/s00348-016-2189-6} {\bibfield  {journal} {\bibinfo
  {journal} {Experiments in Fluids}\ }\textbf {\bibinfo {volume} {57}}
  (\bibinfo {year} {2016}),\ 10.1007/s00348-016-2189-6}\BibitemShut {NoStop}%
\bibitem [{\citenamefont {Wang}\ \emph {et~al.}(2018)\citenamefont {Wang},
  \citenamefont {Gao}, \citenamefont {Wang},\ and\ \citenamefont {Wang}}]{PCS}%
  \BibitemOpen
  \bibfield  {author} {\bibinfo {author} {\bibfnamefont {H.}~\bibnamefont
  {Wang}}, \bibinfo {author} {\bibfnamefont {Q.}~\bibnamefont {Gao}}, \bibinfo
  {author} {\bibfnamefont {Z.}~\bibnamefont {Wang}}, \ and\ \bibinfo {author}
  {\bibfnamefont {J.}~\bibnamefont {Wang}},\ }\bibfield  {title} {\enquote
  {\bibinfo {title} {Error reduction for time‑resolved {PIV} data based on
  {Navier–Stokes} equations},}\ }\href {\doibase 10.1007/s00348-018-2605-1}
  {\bibfield  {journal} {\bibinfo  {journal} {Experiments in Fluids}\ }\textbf
  {\bibinfo {volume} {59}},\ \bibinfo {pages} {149} (\bibinfo {year}
  {2018})}\BibitemShut {NoStop}%
\bibitem [{\citenamefont {Westerweel}(1999)}]{Westerweel1999}%
  \BibitemOpen
  \bibfield  {author} {\bibinfo {author} {\bibfnamefont {J.}~\bibnamefont
  {Westerweel}},\ }\bibfield  {title} {\enquote {\bibinfo {title} {Fundamentals
  of digital particle image velocimetry},}\ }\href {\doibase
  10.1088/0957-0233/8/12/002} {\bibfield  {journal} {\bibinfo  {journal}
  {Measurement Science and Technology}\ }\textbf {\bibinfo {volume} {8}},\
  \bibinfo {pages} {1379} (\bibinfo {year} {1999})}\BibitemShut {NoStop}%
\bibitem [{\citenamefont {Scarano}(2001)}]{Scarano2001}%
  \BibitemOpen
  \bibfield  {author} {\bibinfo {author} {\bibfnamefont {F.}~\bibnamefont
  {Scarano}},\ }\bibfield  {title} {\enquote {\bibinfo {title} {Iterative image
  deformation methods in {PIV}},}\ }\href {\doibase 10.1088/0957-0233/13/1/201}
  {\bibfield  {journal} {\bibinfo  {journal} {Measurement Science and
  Technology}\ }\textbf {\bibinfo {volume} {13}},\ \bibinfo {pages} {R1}
  (\bibinfo {year} {2001})}\BibitemShut {NoStop}%
\bibitem [{\citenamefont {Stanislas}\ \emph {et~al.}(2005)\citenamefont
  {Stanislas}, \citenamefont {Okamoto}, \citenamefont {Kähler},\ and\
  \citenamefont {Westerweel}}]{Stanislas}%
  \BibitemOpen
  \bibfield  {author} {\bibinfo {author} {\bibfnamefont {M.}~\bibnamefont
  {Stanislas}}, \bibinfo {author} {\bibfnamefont {K.}~\bibnamefont {Okamoto}},
  \bibinfo {author} {\bibfnamefont {C.}~\bibnamefont {Kähler}}, \ and\
  \bibinfo {author} {\bibfnamefont {J.}~\bibnamefont {Westerweel}},\ }\bibfield
   {title} {\enquote {\bibinfo {title} {Main results of the second
  international {PIV} challenge},}\ }\href {\doibase 10.1007/s00348-005-0951-2}
  {\bibfield  {journal} {\bibinfo  {journal} {Experiments in Fluids}\ }\textbf
  {\bibinfo {volume} {39}},\ \bibinfo {pages} {170--191} (\bibinfo {year}
  {2005})}\BibitemShut {NoStop}%
\bibitem [{\citenamefont {Horn}\ and\ \citenamefont
  {Schunck}(1981)}]{Horn1981}%
  \BibitemOpen
  \bibfield  {author} {\bibinfo {author} {\bibfnamefont {B.}~\bibnamefont
  {Horn}}\ and\ \bibinfo {author} {\bibfnamefont {B.}~\bibnamefont {Schunck}},\
  }\bibfield  {title} {\enquote {\bibinfo {title} {Determining optical flow},}\
  }\href {\doibase 10.1016/0004-3702(81)90024-2} {\bibfield  {journal}
  {\bibinfo  {journal} {Artificial Intelligence}\ }\textbf {\bibinfo {volume}
  {17}},\ \bibinfo {pages} {185--203} (\bibinfo {year} {1981})}\BibitemShut
  {NoStop}%
\bibitem [{\citenamefont {Corpetti}, \citenamefont {M\'{e}min},\ and\
  \citenamefont {P\'{e}rez}(2002)}]{Corpetti2002}%
  \BibitemOpen
  \bibfield  {author} {\bibinfo {author} {\bibfnamefont {T.}~\bibnamefont
  {Corpetti}}, \bibinfo {author} {\bibfnamefont {E.}~\bibnamefont {M\'{e}min}},
  \ and\ \bibinfo {author} {\bibfnamefont {P.}~\bibnamefont {P\'{e}rez}},\
  }\bibfield  {title} {\enquote {\bibinfo {title} {Dense estimation of fluid
  flows},}\ }\href {\doibase 10.1109/34.990137} {\bibfield  {journal} {\bibinfo
   {journal} {Pattern Analysis and Machine Intelligence, IEEE Transactions on}\
  }\textbf {\bibinfo {volume} {24}},\ \bibinfo {pages} {365--380} (\bibinfo
  {year} {2002})}\BibitemShut {NoStop}%
\bibitem [{\citenamefont {Ruhnau}\ and\ \citenamefont
  {Schnörr}(2007)}]{Ruhnau2007}%
  \BibitemOpen
  \bibfield  {author} {\bibinfo {author} {\bibfnamefont {P.}~\bibnamefont
  {Ruhnau}}\ and\ \bibinfo {author} {\bibfnamefont {C.}~\bibnamefont
  {Schnörr}},\ }\bibfield  {title} {\enquote {\bibinfo {title} {Optical
  {Stokes} flow estimation: an imaging-based control approach},}\ }\href
  {\doibase 10.1007/s00348-006-0220-z} {\bibfield  {journal} {\bibinfo
  {journal} {Experiments in Fluids}\ }\textbf {\bibinfo {volume} {42}},\
  \bibinfo {pages} {61--78} (\bibinfo {year} {2007})}\BibitemShut {NoStop}%
\bibitem [{\citenamefont {Zhong}, \citenamefont {Yang},\ and\ \citenamefont
  {Yin}(2017)}]{Zhong2017}%
  \BibitemOpen
  \bibfield  {author} {\bibinfo {author} {\bibfnamefont {Q.}~\bibnamefont
  {Zhong}}, \bibinfo {author} {\bibfnamefont {H.}~\bibnamefont {Yang}}, \ and\
  \bibinfo {author} {\bibfnamefont {Z.}~\bibnamefont {Yin}},\ }\bibfield
  {title} {\enquote {\bibinfo {title} {An optical flow algorithm based on
  gradient constancy assumption for {PIV} image processing},}\ }\href {\doibase
  10.1088/1361-6501/aa6511} {\bibfield  {journal} {\bibinfo  {journal}
  {Measurement Science and Technology}\ }\textbf {\bibinfo {volume} {28}},\
  \bibinfo {pages} {055208} (\bibinfo {year} {2017})}\BibitemShut {NoStop}%
\bibitem [{\citenamefont {Lu}\ \emph {et~al.}(2021)\citenamefont {Lu},
  \citenamefont {Yang}, \citenamefont {Zhang},\ and\ \citenamefont
  {Yin}}]{lu2021accurate}%
  \BibitemOpen
  \bibfield  {author} {\bibinfo {author} {\bibfnamefont {J.}~\bibnamefont
  {Lu}}, \bibinfo {author} {\bibfnamefont {H.}~\bibnamefont {Yang}}, \bibinfo
  {author} {\bibfnamefont {Q.}~\bibnamefont {Zhang}}, \ and\ \bibinfo {author}
  {\bibfnamefont {Z.}~\bibnamefont {Yin}},\ }\bibfield  {title} {\enquote
  {\bibinfo {title} {An accurate optical flow estimation of {PIV} using fluid
  velocity decomposition},}\ }\href {\doibase 10.1007/s00348-021-03176-w}
  {\bibfield  {journal} {\bibinfo  {journal} {Experiments in Fluids}\ }\textbf
  {\bibinfo {volume} {62}},\ \bibinfo {pages} {1--16} (\bibinfo {year}
  {2021})}\BibitemShut {NoStop}%
\bibitem [{\citenamefont {Cai}\ \emph {et~al.}(2019{\natexlab{b}})\citenamefont
  {Cai}, \citenamefont {Liang}, \citenamefont {Gao}, \citenamefont {Xu},\ and\
  \citenamefont {Wei}}]{cai2019particle}%
  \BibitemOpen
  \bibfield  {author} {\bibinfo {author} {\bibfnamefont {S.}~\bibnamefont
  {Cai}}, \bibinfo {author} {\bibfnamefont {J.}~\bibnamefont {Liang}}, \bibinfo
  {author} {\bibfnamefont {Q.}~\bibnamefont {Gao}}, \bibinfo {author}
  {\bibfnamefont {C.}~\bibnamefont {Xu}}, \ and\ \bibinfo {author}
  {\bibfnamefont {R.}~\bibnamefont {Wei}},\ }\bibfield  {title} {\enquote
  {\bibinfo {title} {Particle image velocimetry based on a deep learning motion
  estimator},}\ }\href {\doibase 10.1109/TIM.2019.2932649} {\bibfield
  {journal} {\bibinfo  {journal} {IEEE Transactions on Instrumentation and
  Measurement}\ }\textbf {\bibinfo {volume} {69}},\ \bibinfo {pages}
  {3538--3554} (\bibinfo {year} {2019}{\natexlab{b}})}\BibitemShut {NoStop}%
\bibitem [{\citenamefont {Schneiders}\ and\ \citenamefont
  {Scarano}(2016)}]{Schneiders2016}%
  \BibitemOpen
  \bibfield  {author} {\bibinfo {author} {\bibfnamefont {J.}~\bibnamefont
  {Schneiders}}\ and\ \bibinfo {author} {\bibfnamefont {F.}~\bibnamefont
  {Scarano}},\ }\bibfield  {title} {\enquote {\bibinfo {title} {Dense velocity
  reconstruction from tomographic {PTV} with material derivatives},}\ }\href
  {\doibase 10.1007/s00348-016-2225-6} {\bibfield  {journal} {\bibinfo
  {journal} {Experiments in Fluids}\ }\textbf {\bibinfo {volume} {57}}
  (\bibinfo {year} {2016}),\ 10.1007/s00348-016-2225-6}\BibitemShut {NoStop}%
\bibitem [{\citenamefont {Schanz}, \citenamefont {Gesemann},\ and\
  \citenamefont {Schröder}(2016)}]{Schanz2016}%
  \BibitemOpen
  \bibfield  {author} {\bibinfo {author} {\bibfnamefont {D.}~\bibnamefont
  {Schanz}}, \bibinfo {author} {\bibfnamefont {S.}~\bibnamefont {Gesemann}}, \
  and\ \bibinfo {author} {\bibfnamefont {A.}~\bibnamefont {Schröder}},\
  }\bibfield  {title} {\enquote {\bibinfo {title} {{Shake-The-Box}:
  {Lagrangian} particle tracking at high particle image densities},}\ }\href
  {\doibase 10.1007/s00348-016-2157-1} {\bibfield  {journal} {\bibinfo
  {journal} {Experiments in Fluids}\ }\textbf {\bibinfo {volume} {57}}
  (\bibinfo {year} {2016}),\ 10.1007/s00348-016-2157-1}\BibitemShut {NoStop}%
\bibitem [{\citenamefont {Wieneke}(2012)}]{Wieneke}%
  \BibitemOpen
  \bibfield  {author} {\bibinfo {author} {\bibfnamefont {B.}~\bibnamefont
  {Wieneke}},\ }\bibfield  {title} {\enquote {\bibinfo {title} {Iterative
  reconstruction of volumetric particle distribution},}\ }\href {\doibase
  10.1088/0957-0233/24/2/024008} {\bibfield  {journal} {\bibinfo  {journal}
  {Measurement Science and Technology}\ }\textbf {\bibinfo {volume} {24}},\
  \bibinfo {pages} {024008} (\bibinfo {year} {2012})}\BibitemShut {NoStop}%
\bibitem [{\citenamefont {Rabault}, \citenamefont {Kolaas},\ and\ \citenamefont
  {Jensen}(2017)}]{rabault2017}%
  \BibitemOpen
  \bibfield  {author} {\bibinfo {author} {\bibfnamefont {J.}~\bibnamefont
  {Rabault}}, \bibinfo {author} {\bibfnamefont {J.}~\bibnamefont {Kolaas}}, \
  and\ \bibinfo {author} {\bibfnamefont {A.}~\bibnamefont {Jensen}},\
  }\bibfield  {title} {\enquote {\bibinfo {title} {Performing particle image
  velocimetry using artificial neural networks: a proof-of-concept},}\ }\href
  {\doibase 10.1088/1361-6501/aa8b87} {\bibfield  {journal} {\bibinfo
  {journal} {Measurement Science and Technology}\ }\textbf {\bibinfo {volume}
  {28}},\ \bibinfo {pages} {125301} (\bibinfo {year} {2017})}\BibitemShut
  {NoStop}%
\bibitem [{\citenamefont {Lee}, \citenamefont {Yang},\ and\ \citenamefont
  {Yin}(2017)}]{lee2017}%
  \BibitemOpen
  \bibfield  {author} {\bibinfo {author} {\bibfnamefont {Y.}~\bibnamefont
  {Lee}}, \bibinfo {author} {\bibfnamefont {H.}~\bibnamefont {Yang}}, \ and\
  \bibinfo {author} {\bibfnamefont {Z.}~\bibnamefont {Yin}},\ }\bibfield
  {title} {\enquote {\bibinfo {title} {{PIV-DCNN}: cascaded deep convolutional
  neural networks for particle image velocimetry},}\ }\href {\doibase
  10.1007/s00348-017-2456-1} {\bibfield  {journal} {\bibinfo  {journal}
  {Experiments in Fluids}\ }\textbf {\bibinfo {volume} {58}},\ \bibinfo {pages}
  {171} (\bibinfo {year} {2017})}\BibitemShut {NoStop}%
\bibitem [{\citenamefont {Wang}\ \emph {et~al.}(2020)\citenamefont {Wang},
  \citenamefont {Yang}, \citenamefont {Li},\ and\ \citenamefont
  {Wang}}]{wang2020}%
  \BibitemOpen
  \bibfield  {author} {\bibinfo {author} {\bibfnamefont {H.}~\bibnamefont
  {Wang}}, \bibinfo {author} {\bibfnamefont {Z.}~\bibnamefont {Yang}}, \bibinfo
  {author} {\bibfnamefont {B.}~\bibnamefont {Li}}, \ and\ \bibinfo {author}
  {\bibfnamefont {S.}~\bibnamefont {Wang}},\ }\bibfield  {title} {\enquote
  {\bibinfo {title} {Predicting the near-wall velocity of wall turbulence using
  a neural network for particle image velocimetry},}\ }\href {\doibase
  10.1063/5.0023786} {\bibfield  {journal} {\bibinfo  {journal} {Physics of
  Fluids}\ }\textbf {\bibinfo {volume} {32}},\ \bibinfo {pages} {115105}
  (\bibinfo {year} {2020})}\BibitemShut {NoStop}%
\bibitem [{\citenamefont {Dosovitskiy}\ \emph {et~al.}(2015)\citenamefont
  {Dosovitskiy}, \citenamefont {Fischer}, \citenamefont {Ilg}, \citenamefont
  {Hausser}, \citenamefont {Hazirbas}, \citenamefont {Golkov}, \citenamefont
  {Van Der~Smagt}, \citenamefont {Cremers},\ and\ \citenamefont
  {Brox}}]{dosovitskiy2015}%
  \BibitemOpen
  \bibfield  {author} {\bibinfo {author} {\bibfnamefont {A.}~\bibnamefont
  {Dosovitskiy}}, \bibinfo {author} {\bibfnamefont {P.}~\bibnamefont
  {Fischer}}, \bibinfo {author} {\bibfnamefont {E.}~\bibnamefont {Ilg}},
  \bibinfo {author} {\bibfnamefont {P.}~\bibnamefont {Hausser}}, \bibinfo
  {author} {\bibfnamefont {C.}~\bibnamefont {Hazirbas}}, \bibinfo {author}
  {\bibfnamefont {V.}~\bibnamefont {Golkov}}, \bibinfo {author} {\bibfnamefont
  {P.}~\bibnamefont {Van Der~Smagt}}, \bibinfo {author} {\bibfnamefont
  {D.}~\bibnamefont {Cremers}}, \ and\ \bibinfo {author} {\bibfnamefont
  {T.}~\bibnamefont {Brox}},\ }\bibfield  {title} {\enquote {\bibinfo {title}
  {Flownet: learning optical flow with convolutional networks},}\ }in\
  \href@noop {} {\emph {\bibinfo {booktitle} {Proceedings of the IEEE
  international conference on computer vision}}}\ (\bibinfo {year} {2015})\
  pp.\ \bibinfo {pages} {2758--2766}\BibitemShut {NoStop}%
\bibitem [{\citenamefont {Ilg}\ \emph {et~al.}(2017)\citenamefont {Ilg},
  \citenamefont {Mayer}, \citenamefont {Saikia}, \citenamefont {Keuper},
  \citenamefont {Dosovitskiy},\ and\ \citenamefont {Brox}}]{Ilg2017}%
  \BibitemOpen
  \bibfield  {author} {\bibinfo {author} {\bibfnamefont {E.}~\bibnamefont
  {Ilg}}, \bibinfo {author} {\bibfnamefont {N.}~\bibnamefont {Mayer}}, \bibinfo
  {author} {\bibfnamefont {T.}~\bibnamefont {Saikia}}, \bibinfo {author}
  {\bibfnamefont {M.}~\bibnamefont {Keuper}}, \bibinfo {author} {\bibfnamefont
  {A.}~\bibnamefont {Dosovitskiy}}, \ and\ \bibinfo {author} {\bibfnamefont
  {T.}~\bibnamefont {Brox}},\ }\bibfield  {title} {\enquote {\bibinfo {title}
  {Flownet 2.0: evolution of optical flow estimation with deep networks},}\ \
  }(\bibinfo {year} {2017})\ pp.\ \bibinfo {pages} {1647--1655}\BibitemShut
  {NoStop}%
\bibitem [{\citenamefont {Sun}\ \emph {et~al.}(2017)\citenamefont {Sun},
  \citenamefont {Yang}, \citenamefont {Liu},\ and\ \citenamefont
  {Kautz}}]{Sun2017}%
  \BibitemOpen
  \bibfield  {author} {\bibinfo {author} {\bibfnamefont {D.}~\bibnamefont
  {Sun}}, \bibinfo {author} {\bibfnamefont {X.}~\bibnamefont {Yang}}, \bibinfo
  {author} {\bibfnamefont {M.-Y.}\ \bibnamefont {Liu}}, \ and\ \bibinfo
  {author} {\bibfnamefont {J.}~\bibnamefont {Kautz}},\ }\bibfield  {title}
  {\enquote {\bibinfo {title} {{PWC-Net}: {CNNs} for optical flow using
  pyramid, warping, and cost volume},}\ }\href {\doibase
  10.1109/CVPR.2018.00931} {\  (\bibinfo {year} {2017}),\
  10.1109/CVPR.2018.00931}\BibitemShut {NoStop}%
\bibitem [{\citenamefont {Hui}, \citenamefont {Tang},\ and\ \citenamefont
  {Loy}(2018)}]{hui2018}%
  \BibitemOpen
  \bibfield  {author} {\bibinfo {author} {\bibfnamefont {T.-W.}\ \bibnamefont
  {Hui}}, \bibinfo {author} {\bibfnamefont {X.}~\bibnamefont {Tang}}, \ and\
  \bibinfo {author} {\bibfnamefont {C.~C.}\ \bibnamefont {Loy}},\ }\bibfield
  {title} {\enquote {\bibinfo {title} {Liteflownet: a lightweight convolutional
  neural network for optical flow estimation},}\ }in\ \href@noop {} {\emph
  {\bibinfo {booktitle} {Proceedings of the IEEE conference on computer vision
  and pattern recognition}}}\ (\bibinfo {year} {2018})\ pp.\ \bibinfo {pages}
  {8981--8989}\BibitemShut {NoStop}%
\bibitem [{\citenamefont {Urolagin}, \citenamefont {kv},\ and\ \citenamefont
  {Reddy}(2011)}]{Urolagin}%
  \BibitemOpen
  \bibfield  {author} {\bibinfo {author} {\bibfnamefont {S.}~\bibnamefont
  {Urolagin}}, \bibinfo {author} {\bibfnamefont {P.}~\bibnamefont {kv}}, \ and\
  \bibinfo {author} {\bibfnamefont {N.~V.~S.}\ \bibnamefont {Reddy}},\
  }\bibfield  {title} {\enquote {\bibinfo {title} {Generalization capability of
  artificial neural network incorporated with pruning method},}\ }in\ \href
  {\doibase 10.1007/978-3-642-29280-4_19} {\emph {\bibinfo {booktitle}
  {International Conference on Advanced Computing, Networking and Security}}},\
  Vol.\ \bibinfo {volume} {7135}\ (\bibinfo {organization} {Springer},\
  \bibinfo {year} {2011})\ pp.\ \bibinfo {pages} {171--178}\BibitemShut
  {NoStop}%
\bibitem [{\citenamefont {Long}, \citenamefont {Shelhamer},\ and\ \citenamefont
  {Darrell}(2015)}]{long2015}%
  \BibitemOpen
  \bibfield  {author} {\bibinfo {author} {\bibfnamefont {J.}~\bibnamefont
  {Long}}, \bibinfo {author} {\bibfnamefont {E.}~\bibnamefont {Shelhamer}}, \
  and\ \bibinfo {author} {\bibfnamefont {T.}~\bibnamefont {Darrell}},\
  }\bibfield  {title} {\enquote {\bibinfo {title} {Fully convolutional networks
  for semantic segmentation},}\ }in\ \href {\doibase 10.1109/CVPR.2015.7298965}
  {\emph {\bibinfo {booktitle} {Proceedings of the IEEE conference on computer
  vision and pattern recognition}}}\ (\bibinfo {year} {2015})\ pp.\ \bibinfo
  {pages} {3431--3440}\BibitemShut {NoStop}%
\bibitem [{\citenamefont {Nie}\ \emph {et~al.}(2018)\citenamefont {Nie},
  \citenamefont {Wang}, \citenamefont {Adeli}, \citenamefont {Lao},
  \citenamefont {Lin},\ and\ \citenamefont {Shen}}]{nie20183}%
  \BibitemOpen
  \bibfield  {author} {\bibinfo {author} {\bibfnamefont {D.}~\bibnamefont
  {Nie}}, \bibinfo {author} {\bibfnamefont {L.}~\bibnamefont {Wang}}, \bibinfo
  {author} {\bibfnamefont {E.}~\bibnamefont {Adeli}}, \bibinfo {author}
  {\bibfnamefont {C.}~\bibnamefont {Lao}}, \bibinfo {author} {\bibfnamefont
  {W.}~\bibnamefont {Lin}}, \ and\ \bibinfo {author} {\bibfnamefont
  {D.}~\bibnamefont {Shen}},\ }\bibfield  {title} {\enquote {\bibinfo {title}
  {{3-D} fully convolutional networks for multimodal isointense infant brain
  image segmentation},}\ }\href {\doibase 10.1109/TCYB.2018.2797905} {\bibfield
   {journal} {\bibinfo  {journal} {IEEE transactions on cybernetics}\ }\textbf
  {\bibinfo {volume} {49}},\ \bibinfo {pages} {1123--1136} (\bibinfo {year}
  {2018})}\BibitemShut {NoStop}%
\bibitem [{\citenamefont {Xu}\ \emph {et~al.}(2015)\citenamefont {Xu},
  \citenamefont {Wang}, \citenamefont {Chen},\ and\ \citenamefont
  {Li}}]{xu2015empirical}%
  \BibitemOpen
  \bibfield  {author} {\bibinfo {author} {\bibfnamefont {B.}~\bibnamefont
  {Xu}}, \bibinfo {author} {\bibfnamefont {N.}~\bibnamefont {Wang}}, \bibinfo
  {author} {\bibfnamefont {T.}~\bibnamefont {Chen}}, \ and\ \bibinfo {author}
  {\bibfnamefont {M.}~\bibnamefont {Li}},\ }\bibfield  {title} {\enquote
  {\bibinfo {title} {Empirical evaluation of rectified activations in
  convolutional network},}\ }\href@noop {} {\bibfield  {journal} {\bibinfo
  {journal} {arXiv preprint arXiv:1505.00853}\ } (\bibinfo {year}
  {2015})}\BibitemShut {NoStop}%
\bibitem [{\citenamefont {Carlier}(2005)}]{carlier2005second}%
  \BibitemOpen
  \bibfield  {author} {\bibinfo {author} {\bibfnamefont {J.}~\bibnamefont
  {Carlier}},\ }\href@noop {} {\enquote {\bibinfo {title} {Second set of fluid
  mechanics image sequences. {European} project fluid image analysis and
  description ({FLUID})},}\ }\bibinfo {howpublished} {\url{http://www.fluid
  .irisa .fr}} (\bibinfo {year} {2005})\BibitemShut {NoStop}%
\bibitem [{\citenamefont {Resseguier}, \citenamefont {M\'{e}min},\ and\
  \citenamefont {Chapron}(2016)}]{Resseguier2016}%
  \BibitemOpen
  \bibfield  {author} {\bibinfo {author} {\bibfnamefont {V.}~\bibnamefont
  {Resseguier}}, \bibinfo {author} {\bibfnamefont {E.}~\bibnamefont
  {M\'{e}min}}, \ and\ \bibinfo {author} {\bibfnamefont {B.}~\bibnamefont
  {Chapron}},\ }\bibfield  {title} {\enquote {\bibinfo {title} {Geophysical
  flows under location uncertainty, part {II} quasi-geostrophy and efficient
  ensemble spreading},}\ }\href {\doibase 10.1080/03091929.2017.1312101}
  {\bibfield  {journal} {\bibinfo  {journal} {Geophysical $\&$ Astrophysical
  Fluid Dynamics}\ }\textbf {\bibinfo {volume} {111}} (\bibinfo {year}
  {2016}),\ 10.1080/03091929.2017.1312101}\BibitemShut {NoStop}%
\bibitem [{\citenamefont {Triantafyllou}, \citenamefont {Triantafyllou},\ and\
  \citenamefont {Grosenbaugh}(1993)}]{Triantafyllou1993}%
  \BibitemOpen
  \bibfield  {author} {\bibinfo {author} {\bibfnamefont {G.}~\bibnamefont
  {Triantafyllou}}, \bibinfo {author} {\bibfnamefont {M.}~\bibnamefont
  {Triantafyllou}}, \ and\ \bibinfo {author} {\bibfnamefont {M.}~\bibnamefont
  {Grosenbaugh}},\ }\bibfield  {title} {\enquote {\bibinfo {title} {Optimal
  thrust development in oscillating foils with application to fish
  propulsion},}\ }\href {\doibase 10.1006/jfls.1993.1012} {\bibfield  {journal}
  {\bibinfo  {journal} {Journal of Fluids and Structures}\ }\textbf {\bibinfo
  {volume} {7}},\ \bibinfo {pages} {205--224} (\bibinfo {year}
  {1993})}\BibitemShut {NoStop}%
\end{thebibliography}%

\end{document}